\documentclass[referee]{raa}            % referee version: for submission
\usepackage{graphicx,times}             %for PS/EPS graphics inclusion, new
\usepackage{natbib}
\usepackage{amssymb,amsmath}
\bibpunct{(}{)}{;}{a}{}{,}

\usepackage[figuresright]{rotating}
\usepackage{tabularx}
\usepackage[pagebackref=true]{hyperref}
\usepackage{threeparttable}

\begin{document}

  \title{Physical Properties of 29 sdB+dM Eclipsing Binaries in Zwicky Transient Facility}
%   \subtitle{I. Place Your Subtitle Here}

   \volnopage{Vol.0 (20xx) No.0, 000--000}      %%preserved for Editor. DOn't remove!
   \setcounter{page}{1}          %%starting page, preserved for Editor. DOn't remove!

   \author{Min Dai  %% Put your Chinese name in "( )" if you like. Note to open line 11 "\usepackage[UTF8]{ctex}"
      \inst{1}
   \and Xiaodian Chen
      \inst{1,2,3}
   \and Kun Wang
      \inst{1}
   \and Yangping Luo
      \inst{1}
   \and Shu Wang
      \inst{1,2}
   \and Licai Deng
      \inst{1,2,3}  
   }

   \institute{School of Physics and Astronomy, China West Normal University, Nanchong 637009, China; {\it chenxiaodian@nao.cas.cn}\\
        \and
             CAS Key Laboratory of Optical Astronomy, National
  Astronomical Observatories, Chinese Academy of Sciences, Beijing
  100101, China\\
        \and
             School of Astronomy and Space Science, University of the Chinese Academy of Sciences, Beijing 101408, China\\
\vs\no
   {\small Received 20xx month day; accepted 20xx month day}}

\abstract{The development of large-scale time-domain surveys provides an opportunity to study the physical properties as well as the evolutionary scenario of B-type subdwarfs (sdB) and M-type dwarfs (dM). Here, we obtained 33 sdB+dM eclipsing binaries based on the Zwicky Transient Facility (ZTF) light curves and {\sl Gaia} early data release 3 (EDR3) parallaxes. By using the PHOEBE code for light curve analysis, we obtain probability distributions for parameters of 29 sdB+dM. $R_1$, $R_2$, and $i$ are well determined, and the average uncertainty of mass ratio $q$ is 0.08. Our parameters are in good agreement with previous works if a typical mass of sdB is assumed. Based on parameters of 29 sdB+dM, we find that both the mass ratio $q$ and the companion's radius $R_2$ decrease with the shortening of the orbital period. For the three sdB+dMs with orbital periods less than 0.075 days, their companions are all brown dwarfs. The masses and radii of the companions satisfy the mass--radius relation for low-mass stars and brown dwarfs. Companions with radii between $0.12R_\odot$ and $0.15R_\odot$ seem to be missing in the observations. As more short-period sdB+dM eclipsing binaries are discovered and classified in the future with ZTF and {\sl Gaia}, we will have more information to constrain the evolutionary ending of sdB+dM.
}

   \authorrunning{M, Dai et al.}            %author_head in even pages
   \titlerunning{sdB+dM Eclipsing Binaries in ZTF}  % title_head in odd pages

   \maketitle
%% The author head (on even pages) and the title head (on odd pages) will be
%% automatically extracted from \author{} and \title{}. Whenever the title is too long,
%% you will be asked to supply a shorter one by inserting either \authorrunning{} or
%% \titlerunning{} before \maketitle. Anyway, you can specify your own heads.
%%
%%
%% Note: In the following text body of your manuscript, please note several differences from
%%       other major journals:
%% (1) \subsection{Please Capitalize the First Letter of Each Notional Word in Subsection Title}
%% (2) Please Capitalize the First Letter of Each Notional Word in all tables' captions

%
%________________________________________________ sections below
%
\section{Introduction}           %% first-level sections will be auto-capitalized
\label{sect:intro}

The hot subdwarf stars are located between the main sequence and the white dwarf (WD) sequence in the Hertzsprung–Russell diagram \citep{Heber_2016}. They are classified as B-type (sdB, $T_{\rm eff}<40000$ K) and O-type (sdO, $T_{\rm eff}>40000$ K) according to their different spectral types \citep{2008ASPC..392..131H}. sdB stars are located in the blue tail of the horizontal branch (HB), which is also known as the extreme horizontal branch \citep[EHB;][]{1986A&A...155...33H}. sdB has a Helium-burning core and an extremely thin residual Hydrogen envelope (mass $<$ 0.01 $M_{\odot}$). Its progenitor is likely an evolved star that loses most of its Hydrogen envelope when it ascends the red giant branch. sdB will not enter the asymptotic giant branch phase, instead, it will become a white dwarf directly. The average lifetime of EHB is in the order of $10^8$ yr \citep{1993ApJ...419..596D}. \citet{2001MNRAS.326.1391M} suggested the binary ratio of sdB is $\sim$ 60$\%$. About half of these binary systems are close binaries with periods of a few days \citep{1994AJ....107.1565A, 2001MNRAS.326.1391M}, while the others are wider binaries with periods of several years \citep{2003AJ....126.1455S,2018MNRAS.473..693V,2013MNRAS.434..186C}.

The evolution of sdB is still controversial \citep{2012A&A...543A.149G}, although the binary evolution scenario is favored. \citet{2002MNRAS.336..449H,2003MNRAS.341..669H} proposed three binary evolution channels for sdBs (common envelope evolution, stable Roche-lobe overflow, and the merger of two He white dwarfs). Subsequently, a number of works tried to testify these theories through observations \citep{2019ApJ...881....7L,2020ApJ...898...64L,2020ApJ...898L..25K,2020A&A...642A..97K,2020A&A...642A.180P}. 
For single sdB stars, the formation has always puzzled us. About $40\%$ of the sdB stars have been found to be single. Although it has been argued that all sdB stars were evolved from binary stars \citep{2020A&A...642A.180P}, but several facts suggest that
the merger channel of two He WDs can not satisfactorily explain the single sdB stars \citep{2020ApJ...905L...7B,2021arXiv210410462V}.
First, the number of low-mass white dwarf binaries is small \citep{2019ApJ...883...51R}. Second, the predicted broad mass distribution of single sdB stars from the merger channel seems to be inconsistent with observations \citep{2012A&A...539A..12F}. Third, observations show that most of the single sdB stars have a very slow rotation, which against the fast rotation predicted by the merger channel \citep{2012A&A...543A.149G,2018OAst...27..112C,2012A&A...543A.149G}. 
Recently, a new channel for the formation of single sdB stars has been proposed \citep{2020ApJ...903..100M}.

When an sdB+dM binary system is eclipsed, it exhibits an HW Vir-type light curve (LC). The number of such stars is relatively small, and since they possesses a characteristic LC, they provide a direct way to measure the physical properties of components by modeling LC and the radial velocity curve, which helps to test the theory of formation and evolution of such systems. For a more detailed review on HW Vir-type stars, see \cite{Heber_2016}. Such stars have been studied continuously and extensively over the last few decades, and their number is increasing (see e.g. \citet{1978MNRAS.183..523K,1986IAUS..118..305M,2001A&A...379..893D,2007ASPC..372..483O,2010ApJ...708..253F,2013MNRAS.430...22B,2014A&A...564A..98S,2015A&A...576A..44K,2017MNRAS.472.3093A,2019PASP..131g5001R,2019MNRAS.490.1283K,2020ApJ...890..126R,2020MNRAS.499.5508S,2021MNRAS.503.3828B}). The orbital periods of HW Vir-type stars are easily obtained from LCs, while the physical parameters such as mass and radius of sdB and the companion stars can be determined based on radial velocity measurements and LC analysis. Based on these parameters, it is possible to constrain the evolution of HW Vir-type stars and the mass--radius relations of two components. The mass--radius relation of brown dwarfs was studied by \citet{2003A&A...402..701B} using the COND model from \citet{2000ApJ...542..464C}. The mass--radius relation of low-mass stars was studied by \citet{2011ApJS..194...28K} using the cataclysmic variables. The HW Vir-type stars are suitable objects for updating the physical properties of low-mass stars and brown dwarfs.

There are currently about 200 HW Vir-type stars in the database of AAVSO International Variable Star Index\footnote{\url{https://www.aavso.org/vsx/index.php}} \citep[VSX]{2006SASS...25...47W}, which are mainly collected from the periodic variable catalog of Zwicky Transient Facility \citep[ZTF]{2020ApJS..249...18C}, the fourth phase of the Optical Gravitational Lensing Experiment \citep[OGLE]{2015AcA....65....1U}, the catalog of the Asteroid Terrestrial-impact Last Alert System \citep[ATLAS]{2018AJ....156..241H}. \citet{2019A&A...630A..80S} studied a sample of HW Vir-type stars in OGLE and ATLAS. ZTF is a northern-sky time-domain survey with limiting magnitudes of 20.8 mag in $g$ band and 20.6 mag in $r$ band \citep{2019PASP..131a8003M}. By monitoring the northern sky with several hundreds of detections for each object over a 3-year period, ZTF will discover a large number of HW Vir-type stars. The presence of more short-period ($P<0.2$ days) HW Vir-type stars will help constrain the evolutionary timescale and ending of such systems.

In this paper, we analyze the physical and orbital parameters of 29 sdB+dM eclipsing binaries using LCs from ZTF data release 5. By fixing sdB's mass and temperature, we obtain the orbital inclination, the mass ratio, the radius of sdB and dM, and the dM temperature. We also determine the mass--radius relation of dM and investigate how sdB+dM evolves with the shortening of the orbital period. In Section 2, we describe the selection of sdB+dM eclipsing binaries and their LCs. We explain how to use PHOEBE to obtain orbital parameter solutions from LCs, and the reliability of our determined parameters in Section 3. In Section 4, we present our results and make a comparison with previous works. We also discuss the physical properties and the evolution stage of our sdB+dM eclipsing binaries in this section. Finally, we conclude this work in Section 5. 

%% Authors can give a citation as 'Michel et al. 1992'.
%% You may also use \cite, \citep and \citet for citation, and use Table~1 or Figure~1
%% and so forth. Using \ref and \label for cross-references of Tables/Figures
%% is a good way in adjusting/adding/removing text, tables or figures.

\section{Sample and Data Selection}
\label{sect:Obs}

\begin{figure}
\centering
\vspace{-0.0in}
\includegraphics[angle=0,width=85mm]{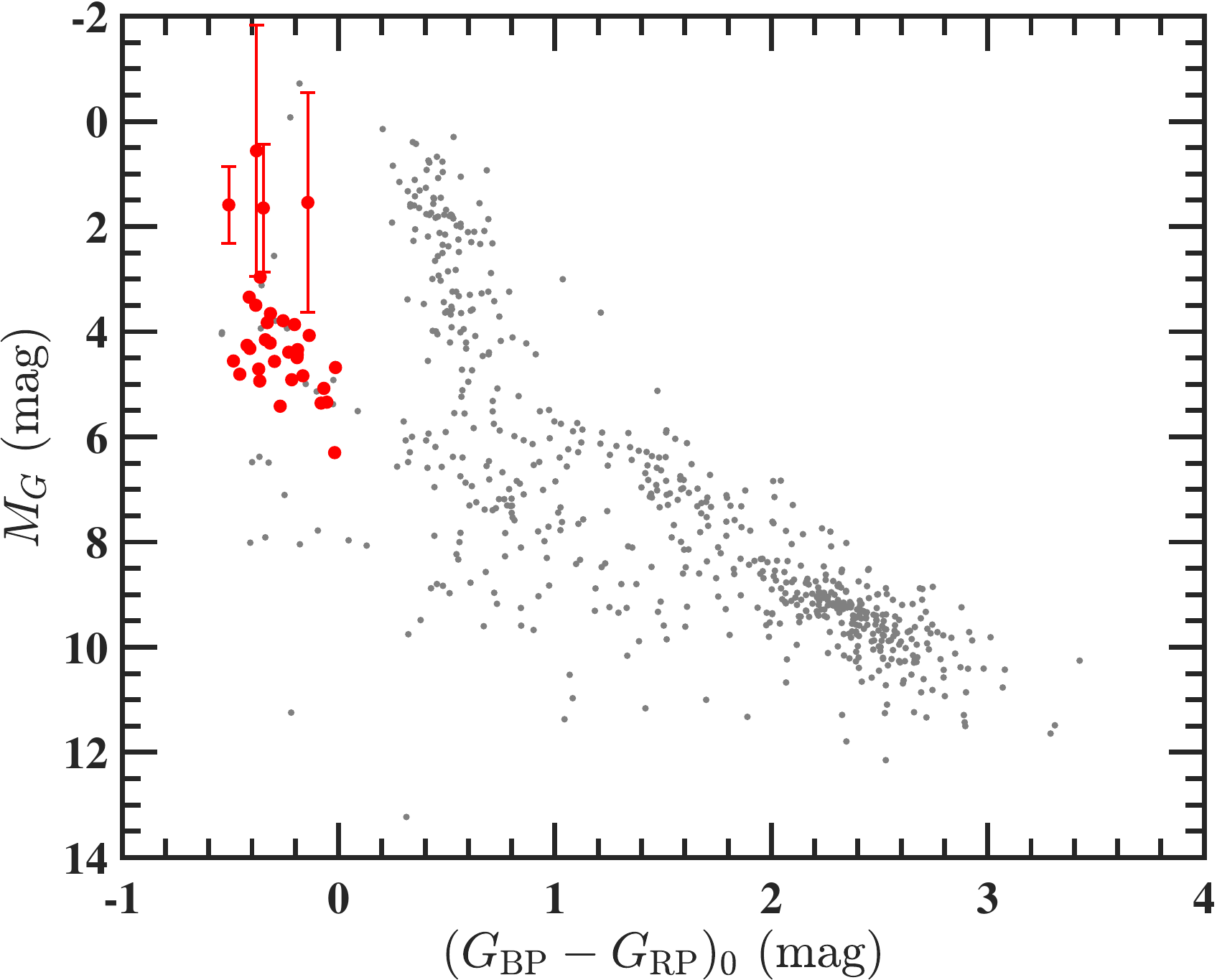}
\vspace{-0.0in}
\caption{{\sl Gaia}-band absolute magnitude vs. intrinsic color diagram. The red dots represent 33 HW Vir-type stars, and the error bars of 4 brighter HW Vir-type stars are added. The gray dots show the locations of ZTF short-period variables with $P<0.2$ days and distance uncertainty less than $50\%$.}
\label{F1}

\end{figure}

We cross-matched the HW Vir-type stars in VSX database with the ZTF catalog of periodic variable stars and obtained a sample of 31 HW Vir-type stars. Two other HW Vir-type stars not included in the VSX were identified by eye and added to the sample. About two-thirds of this sample have periods smaller than 0.2 days. We compared them to other $\sim 900$ short-period variables from ZTF periodic variable catalog through the $\sl Gaia$-band intrinsic color vs. absolute magnitude diagram (CMD). This helps determine the type of primary component of HW Vir-type stars. In Figure~\ref{F1}, the red dots represent the 33 HW Vir-type stars, and the gray dots are the ZTF short-period variables with $P<0.2$ days. We adopted the extinction values of Green's 3D extinction map \citep{2019ApJ...887...93G} and converted them to $A_G$ and $E(G_{\rm Bp}-G_{\rm Rp})$ using the updated extinction law \citep{2019ApJ...877..116W}. The absolute magnitude and intrinsic color are estimated by $M_G=G-A_G-5\log(1/\varpi)-10$ and $(G_{\rm Bp}-G_{\rm Rp})_0=(G_{\rm Bp}-G_{\rm Rp})-E(G_{\rm Bp}-G_{\rm Rp})$, respectively. $\varpi$ is the corrected {\sl Gaia} early data release 3 (EDR3) parallax in $m$as \citep{2021A&A...649A...4L}. The corrected EDR3 parallax was found to reduce the systematic bias to $<10$ $\mu$as \citep{2021ApJ...911L..20R}.

From Figure \ref{F1}, we found that HW Vir-type stars concentrate as a clump with an absolute magnitude around $M_G=5$ mag. The clump is bluer than the main-sequence variables and brighter than the cataclysmic variables. This means our HW Vir-type stars are sdB+dM eclipsing binaries rather than WD+dM eclipsing binaries. To further confirm that primary components our sample are all sdB stars, we adopt the method in \citep{2015MNRAS.448.2260G}, where the authors calculated the reduced proper motion defined as $H$ = $G$ + 5 log$\mu$ + 5. $G$ and $\mu$ are $G$-band magnitude and proper motion. We found that $H$ of our sample are all less than 14.5, which proves that they are sdBs rather than WDs \citep{2019A&A...630A..80S}. The distribution of HW Vir-type stars is consistent with constraints of $-1.0< M_{G} < 7.0$ and $(G_{BP}-G_{RP})_0<0.7$ \citep{2019A&A...621A..38G, 2020A&A...635A.193G}, but more concentrated in $M_{G}$. Four sdB+dM are brighter than the others, and these deviations are due to the large uncertainties propagated from {\sl Gaia} parallaxes. We showed their error bars in Figure~\ref{F1}. The $1\sigma$ uncertainties of the absolute magnitude for other sdB+dM are less than 0.35 mag. We also noted that in CMD, there are dozens of gray dots distributed in the location of our sdB+dM eclipsing binaries. By examining their LCs, we found that they are non-eclipsing sdB+dM binaries, and their brightness variations are due to reflection. Reflection occurs when the hot component illuminates part of the spherical surface of the cooler component when the temperature difference between the two components is large. The size of the illuminated surface varies with the projection angle, which leads to an eventual sinusoidal-like light curve without eclipse. This suggests that many sdB binaries, which orbit at such a small inclination that we cannot see the eclipse. The HW Vir-type star is a subtype of detached eclipsing binaries (EA-type) that have similar LC shapes. The specific distribution in the intrinsic color vs. absolute magnitude diagram helps to separate HW Vir-type stars from other main-sequence eclipsing binaries. With these properties, we expect to find hundreds of HW Vir-type stars in future ZTF-based variable star searches. Besides, CMD is useful to separate sdB+dM eclipsing binaries and WD+dM eclipsing binaries when {\sl Gaia} parallaxes are accurate enough.

%then we visually checked these LC  
We downloaded the $g,r$-band LCs of 33 sdB+dM eclipsing binaries from the ZTF DR5 database. To ensure the quality of LCs, we selected only good photometric data with ‘catflag $<$ 10’. ZTF DR5 contains photometric data obtained during a two-and-a-half year survey, with more than 200 detections per band for most objects. Given that typical photometric uncertainties are around 0.015 mag, we believed that a reliable analysis can be performed on the ZTF LCs. LCs were folded with periods from \citet{2020ApJS..249...18C}, and the ephemeris of the primary minimum $T_{0}$ was estimated by lc\_geometry package in PHOEBE 2.3 \citep{2020ApJS..250...34C}. We also performed a careful visual inspection of all light curves to make sure that the period and $T_{0}$ were not significantly biased. We excluded photometric outliers (less than 10) for each LC according to the fit line of the Gaussian processes. Magnitude was converted into normalized flux by assuming the maximum flux in $g,r$ bands is equal to 1. The primary eclipse of four sdB+dM is fainter than the detection limit of ZTF ($r>20.5$ mag). We excluded them and performed LC analysis on the remaining 29 sdB+dM.

\section{light curve analysis}
\label{sect:data}

sdB+dM eclipsing binaries are the best objects to study the physical properties of sdB and dM. However, only the radial velocity curve of the primary component is available, so the mass ratio cannot be directly determined from radial velocity curves. One way to solve this problem is to assume a typical mass of sdB, e.g., $M_1=0.47M_\odot$. Based on this assumption, parameters of two components can be well constrained by the analysis of the single-line radial velocity curve and light curves. The shortcoming of this method is that $M_1$ cannot be studied. The other way is to constrain the mass ratio $q$ based on $\log g$ and fix it in the subsequent analysis \citep{2010ApJ...708..253F,2021MNRAS.501.3847S,2001A&A...379..893D,2004A&A...420..251H,2007A&A...471..605V,2014A&A...564A..98S,2017MNRAS.472.3093A,2013MNRAS.430...22B}. In this method, the accuracy of $q$ depends on the extent to which it is constrained by LCs. When $q$ does not converge in the iterations, the uncertainty is relatively larger. In this work, we used the PHOEBE 2.3 code to study the probability distribution of parameters for 29 sdB+dM and to evaluate how well $q$ is constrained by LCs.

With some exceptions, most sdB stars have a mass around 0.47 $M_{\odot}$, which is also known as the typical mass \citep{2002MNRAS.336..449H,2003MNRAS.341..669H,2020MNRAS.495.2844S}. Most sdB stars have a temperature in range of $27,000 - 31,000$ K \citep{2012ASPC..452...57G}. 3 of 29 sdB+dM were studied previously with spectra and sdB's temperature is in this range. The three sdB+dM are ZTF J162256.66+473051.1 with a temperature $T_{1, \rm eff} = 29000 \pm 600$ K \citep{2014A&A...564A..98S}, ZTF J153349.44+375927.8 with a temperature $T_{1, \rm eff} = 29230 \pm 125$ K \citep{2010ApJ...708..253F}, and ZTF J223421.49+245657.1 with a temperature $T_{1, \rm eff} = 28500 \pm 500$ K \citep{2017MNRAS.472.3093A} or $T_{1, \rm eff} = 28370 \pm 80 $ K \citep{2007ASPC..372..483O}. The companion star of sdB+dM is an M dwarf or a brown dwarf, and its temperature is around 3000 K \citep{1986IAUS..118..305M,2014A&A...564A..98S,2014ASPC..481..259V}.

Our idea is to assume that the primary components of sdB+dM are all canonical sdB star with a mass of 0.47 $\pm \  0.03$ $M_{\odot}$, a radius of 0.175 $\pm \  0.025$ $R_{\odot}$, and temperature about 30,000 K \citep{2021arXiv210410462V}. We fixed the sdB mass $M_{1}$ to 0.47 $M_{\odot}$, the sdB temperature $T_{1}$ to 30,000 K in our LC analysis. Small deviations in $T_{1}$ will directly affect the poorly constrained $T_{2}$ (typically with $30\%$ uncertainty), but have little effect on $R_2$ and other parameters. With these assumptions, we can still study the properties of the companion star and the evolution stage of the system in statistics. We set the orbital eccentricity $e$ to 0 since all of our sdB+dM eclipsing binaries have a very short period (less than 0.5 days). We set the linear limb-darkening coefficient of sdB to 0.25 and 0.2 in the $g$ band and $r$ band, respectively, while the companion is 1.0 in both the $g$ band and $r$ band \citep{2011A&A...529A..75C}. we set the gravity-darkening factor to 1.0 and 0.32 for sdB and the companion, respectively \citep{2011A&A...529A..75C}. The reflection factor of sdB and the companion was set to 1.0.  The range of the initial orbital inclination $i$ is $60-90$ degrees, and the other parameters are unrestricted. All preset parameters are list in Table \ref{T1}.

\begin{table}

\begin{center}
   \begin{threeparttable}[b]
\caption[]{Preset MCMC parameters in PHOEBE 2.3}\label{T1}

\begin{tabular}{lc}

\hline\noalign{\smallskip}

\multicolumn{2}{c}{Fixed parameters:}\\
  \hline\noalign{\smallskip}
$M_{1}$&0.47 $\rm M_{\odot}$\\
$T_{1}$&30000 K\\
ecc&0\\
$x_{1r}$\tnote{1}&0.2\\
$x_{1g}$\tnote{1}&0.25\\
$x_{2r}$\tnote{1}&1.0\\
$x_{2g}$\tnote{1}&1.0\\
$g_{1}$\tnote{1}&1.0\\
$g_{2}$\tnote{1}&0.32\\
$A_{1}$\tnote{2}&1.0\\
$A_{2}$\tnote{2}&1.0\\
  \hline\noalign{\smallskip}
\multicolumn{2}{c}{fitting with MCMC Calculation parameters:}\\
  \hline\noalign{\smallskip}
$i$ &$60 - 90 ^{\circ}$\\
$q (= M_{2}/M_{1})$&unrestricted\\
$R_{1}$&unrestricted\\
$R_{2}$&unrestricted\\
$T_{2}$&unrestricted\\
$L_{1}$&unrestricted\\
$L_{2}$&unrestricted\\
  \hline\noalign{\smallskip}
\end{tabular}
\begin{tablenotes}
     \item[1] Linear limb darkening coefficients and gravity-darkening factors adopted from \citep{2011A&A...529A..75C}.
     \item[2] reflection factors.
   \end{tablenotes}
  \end{threeparttable}
\end{center}
\end{table}

We used the Markov chain Monte Carlo (MCMC) sampler based on EMCEE \citep{2013PASP..125..306F} to determine parameters of 29 sdB+dM eclipsing binaries. For sampling, we used 50 walkers and 2000 iterations, and each run costs 30 hours on a 4-core CPU. We also tested 5000 iterations for 5 sdB+dM and 20,000 iterations for one sdB+dM and found the results were almost the same as those based on 2000 iterations. In the first run, the mass ratios $q_i$ are not well constrained for most of the objects, as it is difficult to obtain the true $q$ from many local optimal mass ratios. In contrast, the radii $R_{1,i}, R_{2,i}$  and inclination $i_i$ are well constrained in two-band LC analysis. From the $R_{1,i}-q$ and $R_{2,i}-q$ distributions, $R_{1,i}, R_{2,i}$ only increase by $20\%$ when $q$ increases from 0 to 1. So we decided to use $R_{2,i}$ to establish a prior distribution of $q$. According to the mass--radius relationship of dM, the ratio of mass to radius is slightly less than 1 in the solar unit. Given that $q$ and $R_{2}$ are usually overestimated in the first run ($q$ of the sdB+dM is usually less than 0.3), we assumed a prior distribution of $q$ as [0, $R_{2,i}$/0.47]. We again performed MCMC estimation and obtained the reliable parameter distributions for most of sdB+dM. 11 sdB+dM eclipsing binaries have temperatures $T_{2}>>3000$ K are not consistent with their masses and radii. For these objects, we added another prior $T_2<4000$ K and performed the MCMC estimation again to obtain the final distributions of parameters. 

\begin{figure}[h]
  \begin{minipage}[t]{0.48\linewidth}
  \centering
   \includegraphics[width=\linewidth]{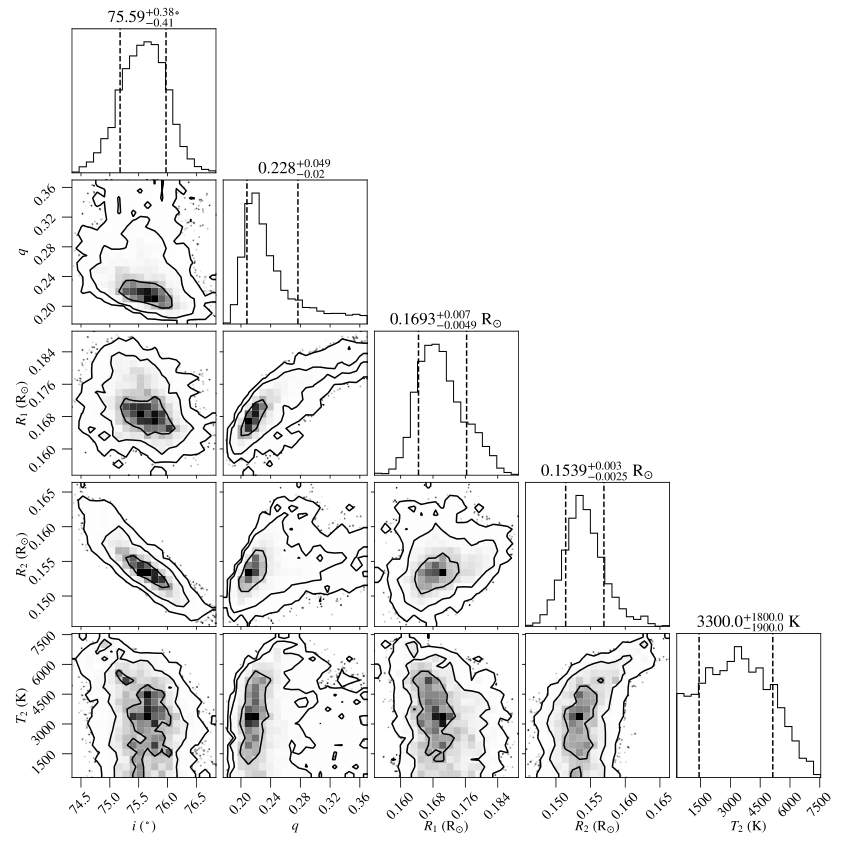}
	  \caption{\label{F2}{\small MCMC results for ZTF J204046.56+340702.8. In the diagram, $i$, $R_{1}$, $R_{2}$, and $q$ present a Gaussian-like distribution.} }
  \end{minipage}%
  \hspace{0.04\linewidth}
  \begin{minipage}[t]{0.48\linewidth}
  \centering
   \includegraphics[width=\linewidth]{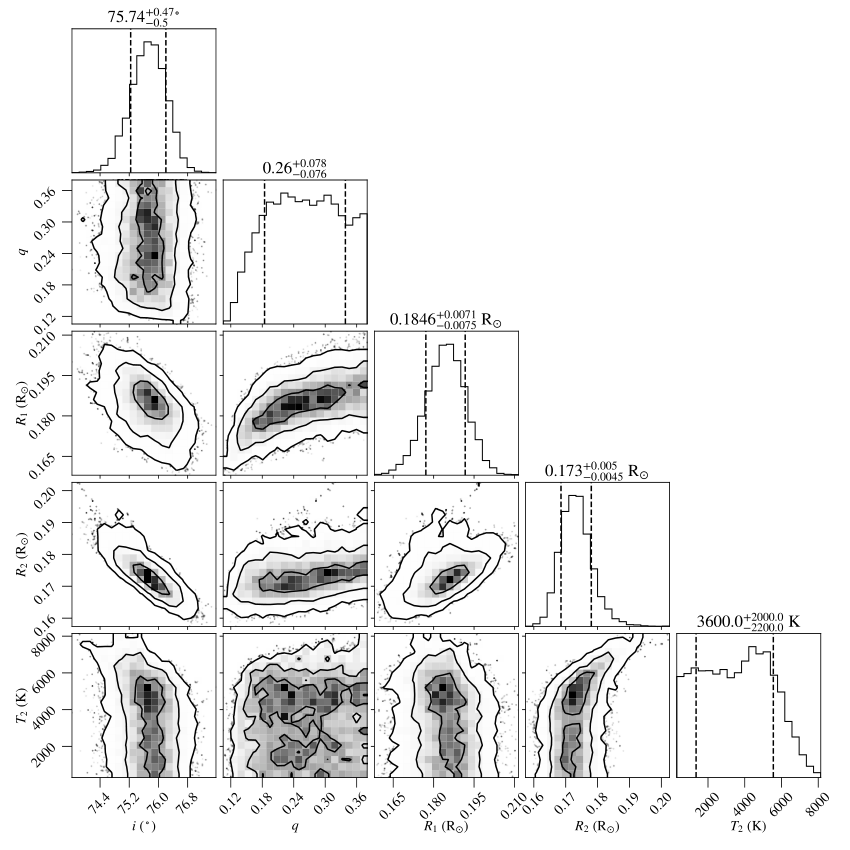}
	  \caption{\label{F3}{\small MCMC results for ZTF J224547.36+490824.7. In the diagram, $i$, $R_{1}$, and $R_{2}$ present a Gaussian-like distribution, but $q$ does not obey a Gaussian distribution.}}
  \end{minipage}%
\end{figure}

Figure \ref{F2} shows the parameter distributions of ZTF J204046.56+340702.8 as an example of one class. We can see that $i$, $q$, $R_{1}$ and $R_{2}$ obey Gaussian-like distributions and the correlations between these parameters are small. Parameters of the other three sdB+dM eclipsing binaries are similarly distributed (See Figures in Appendix A). Figure \ref{F3} shows the parameter distributions of ZTF J224547.36+490824.7. It differs from Figure \ref{F2} in that $q$ does not satisfy the Gaussian distribution. For this class, $q$ is uniformly distributed, or has a wide tail, or is biased towards one side in the prior range (See Figures in Appendix B). Considering that the prior range is narrow, i.e., $q_i$ is roughly distributed between 0 and $0.2-0.5$, the deviation between the maximum likelihood and the true $q$ is not large. This deviation is already included in the error of $q$ (mean uncertainty is 0.08). Besides, the error in $R_2$ hardly increases with the error in $q$ due to the weak correlation between $R_2$ and $q$. Based on multi-band LC analysis, the determined masses and radii of the companions in sdB+dM eclipsing binaries are reliable for studying their statistical properties. Nevertheless, in most cases, uncertainty in $q$ or $M_2$ is larger.

\begin{figure*}[h]
  \begin{minipage}[t]{0.48\linewidth}
  \centering
   \includegraphics[width=\linewidth]{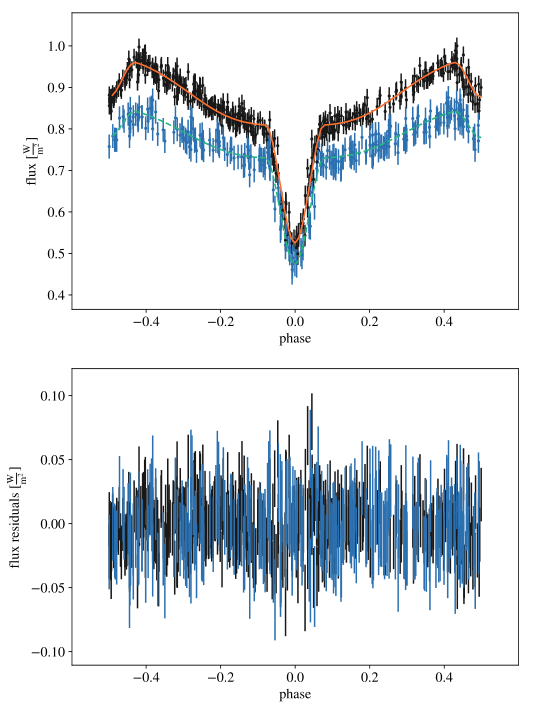}
	  \caption{\label{F4}{\small LC analysis of ZTF J204046.56+340702.8. The top and bottom panels are the LC fitting diagram and residual diagram. In the top panel, the green and yellow lines are modelled LCs, and the blue and black dots are photometric measurements in $g$ and $r$ band. The error bars show the uncertainties of the photometry. In the bottom panel, the blue and black dots are the residuals between the modelled and observed LCs. The green and yellow lines indicate the average of the residuals.} }
  \end{minipage}%
  \hspace{0.04\linewidth}
  \begin{minipage}[t]{0.48\linewidth}
  \centering
   \includegraphics[width=\linewidth]{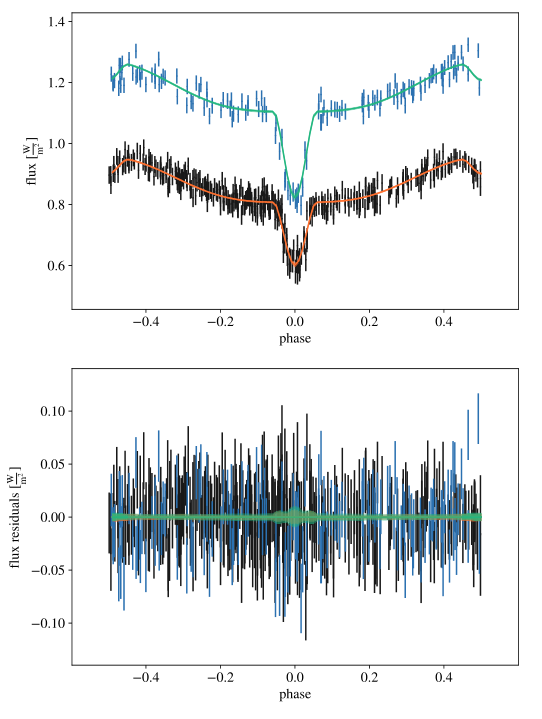}
	  \caption{\label{F5}{\small Similar to Figure \ref{F4}, but for ZTF J224547.36+490824.7.}}
  \end{minipage}%
\end{figure*}

In both Figure \ref{F2} and Figure \ref{F3}, $T_2$ does not show a Gaussian distribution, which means that $T_2$ is difficult to be constrained in the light curve analysis. The typical uncertainty of $1500$ K also indicates that $T_2$ is rather uncertain. The upper panels of Figures \ref{F4} and \ref{F5} show a comparison of the modeled and observed light curves of ZTF J204046.56+340702.8 and ZTF J224547.36+490824.7, while the lower panels show the residual diagrams. Diagrams for other sdB+dM eclipsing binaries are supplemented in Appendix A and B.

\section{Results and discussions}

In this section, we present a table including parameters of 29 sdB+dM eclipsing binaries and make a comparison with previous works. We also discuss how to use our result to constrain the evolutionary stage of sdB+dM eclipsing binaries and the physical properties of companion dM stars.

\subsection{results of 29 sdB+dM eclipsing binaries}
The results of our sample, containing 29 sdB+dM, were included in Table \ref{T2}. It contains the ZTF ID, position (J2000 R.A. and decl.), period, physical parameters and the corresponding uncertainties. The physical parameters include $i$, $q$, $R_{1}$, $R_{2}$, $T_{2}$, $L_{1}$, and $L_{2}$. The bolometric luminosities of sdB and companion are calculated using the PHOEBE code, converted from the passband luminosities. The uncertainty of luminosity is simply estimated based on the propagation uncertainties of the radius and temperature through the equation $L=4\pi \sigma R^2 T_{\rm eff}^4$. The mean radius and luminosity of sdB are $R_1=0.188\pm0.018M_\odot$ and $L_1=21\pm4L_\odot$, which are consistent with the typical radius and luminosity of sdB \citep{Heber_2016}. 
%\iffalse

\setlength{\tabcolsep}{2pt}
\begin{table*}

\tiny

\begin{center}
\caption[]{Physical parameters of 29 sdB+dM}\label{T2}

\centering

\begin{tabular}{ccccccccccc}

\hline\noalign{\smallskip}
ID&R.A. (J2000)&Decl. (J2000) & Period & $i$ & $q$ & $R_{1} $ & $R_{2}$ & $T_{2}$&$L_{1}$& $L_{2}$\\     
    & (deg) & (deg) & (days) & ($^\circ$) & &($ R_{\odot}$) & ($R_{\odot})$ & (K) & ($L_{\odot}$) & ($ L_{\odot}$)\\

\hline\noalign{\smallskip}

ZTF J011339.09+225739.0&18.41288&22.96086&0.093373&$79.6^{+0.2}_{-0.3}$&$0.29^{+0.06}_{-0.09}$&$0.178^{+0.004}_{-0.007}$&$0.160^{+0.002}_{-0.002}$&$3200^{+1600}_{-1900}$&$18^{+1}_{-2}$&$0.002^{+0.005}_{-0.006}$\\
ZTF J014600.90+581420.4&26.50378&58.23902&0.0937295&$88.2^{+1.2}_{-1.3}$&$0.28^{+0.16}_{-0.08}$&$0.159^{+0.008}_{-0.009}$&$0.167^{+0.004}_{-0.004}$&$2600^{+1900}_{-1600}$&$15^{+2}_{-2}$&$0.001^{+0.003}_{-0.003}$\\
ZTF J050114.39+424741.3&75.30996&42.79481&0.1385308&$78.7^{+0.3}_{-0.3}$&$0.24^{+0.15}_{-0.03}$&$0.203^{+0.011}_{-0.007}$&$0.220^{+0.004}_{-0.003}$&$2600^{+1100}_{-1600}$&$24^{+3}_{-2}$&$0.001^{+0.003}_{-0.005}$\\
ZTF J054744.02+304732.2&86.93345&30.79228&0.0946493&$79.8^{+0.9}_{-1.0}$&$0.35^{+0.10}_{-0.09}$&$0.190^{+0.009}_{-0.010}$&$0.189^{+0.007}_{-0.007}$&$2300^{+1200}_{-1400}$&$21^{+2}_{-3}$&$0.001^{+0.002}_{-0.002}$\\
ZTF J072905.44-183703.4&112.27271&-18.61767&0.0937585&$65.4^{+0.7}_{-0.7}$&$0.41^{+0.04}_{-0.03}$&$0.204^{+0.008}_{-0.008}$&$0.223^{+0.005}_{-0.005}$&$2500^{+1500}_{-1500}$&$24^{+2}_{-2}$&$0.002^{+0.004}_{-0.004}$\\
ZTF J153349.44+375927.8&233.45600&37.99114&0.1617705&$86.7^{+0.1}_{-0.1}$&$0.37^{+0.02}_{-0.05}$&$0.177^{+0.001}_{-0.002}$&$0.165^{+0.001}_{-0.002}$&$2700^{+800}_{-1400}$&$18^{+0.3}_{-1}$&$0.001^{+0.002}_{-0.003}$\\
ZTF J162256.66+473051.1&245.73609&47.51422&0.0697888&$69.4^{+0.5}_{-0.5}$&$0.13^{+0.04}_{-0.03}$&$0.182^{+0.004}_{-0.004}$&$0.098^{+0.002}_{-0.002}$&$2700^{+1000}_{-1500}$&$17.0^{+1.0}_{-1.0}$&$0.0003^{+0.0007}_{-0.001}$\\
ZTF J183431.88+061056.7&278.63286&6.18243&0.0612955&$80.4^{+0.6}_{-0.6}$&$0.17^{+0.04}_{-0.04}$&$0.148^{+0.004}_{-0.004}$&$0.101^{+0.003}_{-0.003}$&$2300^{+1200}_{-1400}$&$13^{+1}_{-1}$&$0.0002^{+0.0006}_{-0.0006}$\\
ZTF J183522.73+064247.1&278.84475&6.71306&0.1544056&$83.9^{+0.4}_{-0.4}$&$0.41^{+0.09}_{-0.13}$&$0.160^{+0.006}_{-0.007}$&$0.198^{+0.005}_{-0.006}$&$2900^{+1600}_{-1700}$&$15^{+1}_{-2}$&$0.002^{+0.006}_{-0.006}$\\
ZTF J184042.41+070321.9&280.17675&7.056056&0.1911135&$77.1^{+0.3}_{-0.3}$&$0.16^{+0.02}_{-0.01}$&$0.194^{+0.007}_{-0.009}$&$0.246^{+0.004}_{-0.004}$&$3600^{+300}_{-1500}$&$22^{+2}_{-3}$&$0.005^{+0.004}_{-0.015}$\\
ZTF J184847.05+115720.3&282.19608&11.95565&0.1068316&$87.8^{+0.6}_{-0.6}$&$0.35^{+0.07}_{-0.06}$&$0.194^{+0.005}_{-0.006}$&$0.201^{+0.002}_{-0.002}$&$2300^{+1300}_{-1300}$&$22.0^{+1.0}_{-2.0}$&$0.001^{+0.002}_{-0.002}$\\
ZTF J185207.60+144547.1&283.03171&14.76314&0.190421&$83.7^{+0.1}_{-0.1}$&$0.53^{+0.04}_{-0.10}$&$0.187^{+0.003}_{-0.004}$&$0.194^{+0.002}_{-0.004}$&$3600^{+300}_{-900}$&$20^{+1}_{-1}$&$0.004^{+0.002}_{-0.006}$\\
ZTF J190705.22+323216.9&286.77179&32.53803&0.1380584&$80.7^{+0.5}_{-0.5}$&$0.26^{+0.03}_{-0.06}$&$0.204^{+0.007}_{-0.007}$&$0.112^{+0.004}_{-0.004}$&$2300^{+1200}_{-1300}$&$24^{+2}_{-2}$&$0.0003^{+0.0007}_{-0.0007}$\\
ZTF J192055.46+041619.5&290.23113&4.27208&0.1117401&$76.0^{+0.8}_{-0.8}$&$0.5^{+0.06}_{-0.05}$&$0.183^{+0.011}_{-0.009}$&$0.273^{+0.010}_{-0.009}$&$2800^{+1600}_{-1700}$&$19^{+3}_{-2}$&$0.004^{+0.010}_{-0.010}$\\
ZTF J192240.88+262415.5&290.67034&26.40433&0.1519447&$87.9^{+1.3}_{-1.0}$&$0.17^{+0.23}_{-0.08}$&$0.200^{+0.012}_{-0.012}$&$0.178^{+0.004}_{-0.004}$&$2800^{+1400}_{-1700}$&$23^{+3}_{-3}$&$0.001^{+0.004}_{-0.004}$\\
ZTF J192513.66+253025.6&291.30696&25.50711&0.1720741&$83.8^{+0.7}_{-0.7}$&$0.31^{+0.17}_{-0.18}$&$0.179^{+0.013}_{-0.015}$&$0.201^{+0.008}_{-0.008}$&$2200^{+1200}_{-1300}$&$18^{+3}_{-4}$&$0.001^{+0.002}_{-0.002}$\\
ZTF J193555.33+123754.8&293.98056&12.63191&0.2458299&$80.0^{+0.3}_{-0.3}$&$0.12^{+0.02}_{-0.01}$&$0.181^{+0.008}_{-0.007}$&$0.272^{+0.005}_{-0.005}$&$2500^{+1200}_{-1500}$&$19^{+2}_{-2}$&$0.002^{+0.005}_{-0.006}$\\
ZTF J193604.87+371017.2&294.02038&37.1715&0.2383538&$82.1^{+0.3}_{-0.3}$&$0.27^{+0.13}_{-0.20}$&$0.209^{+0.010}_{-0.014}$&$0.189^{+0.006}_{-0.006}$&$3000^{+1700}_{-1800}$&$25^{+3}_{-4}$&$0.002^{+0.006}_{-0.006}$\\
ZTF J193737.06+092638.7&294.40442&9.44408&0.1741799&$83.5^{+0.4}_{-0.4}$&$0.29^{+0.26}_{-0.09}$&$0.180^{+0.014}_{-0.009}$&$0.249^{+0.009}_{-0.005}$&$2200^{+1300}_{-1300}$&$19^{+4}_{-2}$&$0.001^{+0.003}_{-0.003}$\\
ZTF J195403.63+355700.6&298.51514&35.95018&0.0626614&$81.5^{+1.3}_{-1.2}$&$0.23^{+0.04}_{-0.05}$&$0.205^{+0.004}_{-0.004}$&$0.106^{+0.003}_{-0.002}$&$3700^{+1000}_{-1900}$&$24^{+1}_{-1}$&$0.001^{+0.002}_{-0.004}$\\
ZTF J195908.44+365041.1&299.78523&36.84477&0.476099&$85.8^{+0.1}_{-0.1}$&$0.37^{+0.06}_{-0.07}$&$0.211^{+0.005}_{-0.006}$&$0.237^{+0.004}_{-0.004}$&$1500^{+900}_{-900}$&$26^{+2}_{-2}$&$0.0002^{+0.0007}_{-0.0006}$\\
ZTF J200411.60+141150.2&301.04833&14.19728&0.2564331&$84.2^{+0.1}_{-0.2}$&$0.29^{+0.16}_{-0.14}$&$0.182^{+0.008}_{-0.009}$&$0.241^{+0.008}_{-0.005}$&$2200^{+1300}_{-1300}$&$19^{+2}_{-2}$&$0.001^{+0.003}_{-0.003}$\\
ZTF J203535.01+354405.0&308.89596&35.73475&0.2044668&$87.2^{+0.3}_{-0.3}$&$0.54^{+0.05}_{-0.21}$&$0.220^{+0.003}_{-0.010}$&$0.221^{+0.003}_{-0.010}$&$3000^{+1100}_{-1700}$&$27^{+1}_{-3}$&$0.003^{+0.005}_{-0.009}$\\
ZTF J204046.56+340702.8&310.19402&34.11746&0.0746827&$75.6^{+0.4}_{-0.4}$&$0.23^{+0.05}_{-0.02}$&$0.169^{+0.007}_{-0.005}$&$0.154^{+0.003}_{-0.003}$&$3300^{+1800}_{-1900}$&$17.0^{+2.0}_{-1.0}$&$0.002^{+0.006}_{-0.006}$\\
ZTF J204638.16+514735.5&311.65904&51.79322&0.0896432&$63.6^{+0.2}_{-0.1}$&$0.55^{+0.002}_{-0.005}$&$0.163^{+0.001}_{-0.001}$&$0.245^{+0.001}_{-0.001}$&$2800^{+800}_{-1400}$&$15^{+0.2}_{-0.2}$&$0.002^{+0.004}_{-0.007}$\\
ZTF J210401.41+343636.3&316.00592&34.61008&0.1185524&$80.5^{+0.2}_{-0.2}$&$0.35^{+0.14}_{-0.06}$&$0.197^{+0.009}_{-0.007}$&$0.219^{+0.004}_{-0.003}$&$2000^{+1200}_{-1200}$&$23^{+3}_{-2}$&$0.001^{+0.002}_{-0.002}$\\
ZTF J221339.18+445155.8&333.41329&44.86552&0.2536091&$83.2^{+0.4}_{-0.5}$&$0.12^{+0.29}_{-0.08}$&$0.204^{+0.017}_{-0.014}$&$0.183^{+0.007}_{-0.007}$&$3300^{+1900}_{-2000}$&$24^{+5}_{-4}$&$0.004^{+0.009}_{-0.009}$\\
ZTF J223421.49+245657.1&338.58950&24.94928&0.1105878&$78.6^{+0.4}_{-0.4}$&$0.2^{+0.03}_{-0.06}$&$0.204^{+0.004}_{-0.006}$&$0.108^{+0.003}_{-0.003}$&$3400^{+2100}_{-2100}$&$20^{+1}_{-2}$&$0.001^{+0.003}_{-0.003}$\\
ZTF J224547.36+490824.7&341.44742&49.14019&0.1207436&$75.7^{+0.5}_{-0.5}$&$0.26^{+0.08}_{-0.08}$&$0.185^{+0.007}_{-0.008}$&$0.173^{+0.005}_{-0.005}$&$3600^{+2000}_{-2200}$&$20^{+2}_{-2}$&$0.004^{+0.010}_{-0.011}$\\
\hline\noalign{\smallskip}
\end{tabular}

%\tablenotetext{a}{The entire table is available in the online journal;
 % 50 lines are shown here for guidance regarding its form and
 % content.}
\end{center}
\end{table*}

%\fi 

\subsection{External errors of orbital parameters}

\begin{figure*}
\centering
\vspace{-0.0in}
\includegraphics[angle=0,width=\linewidth]{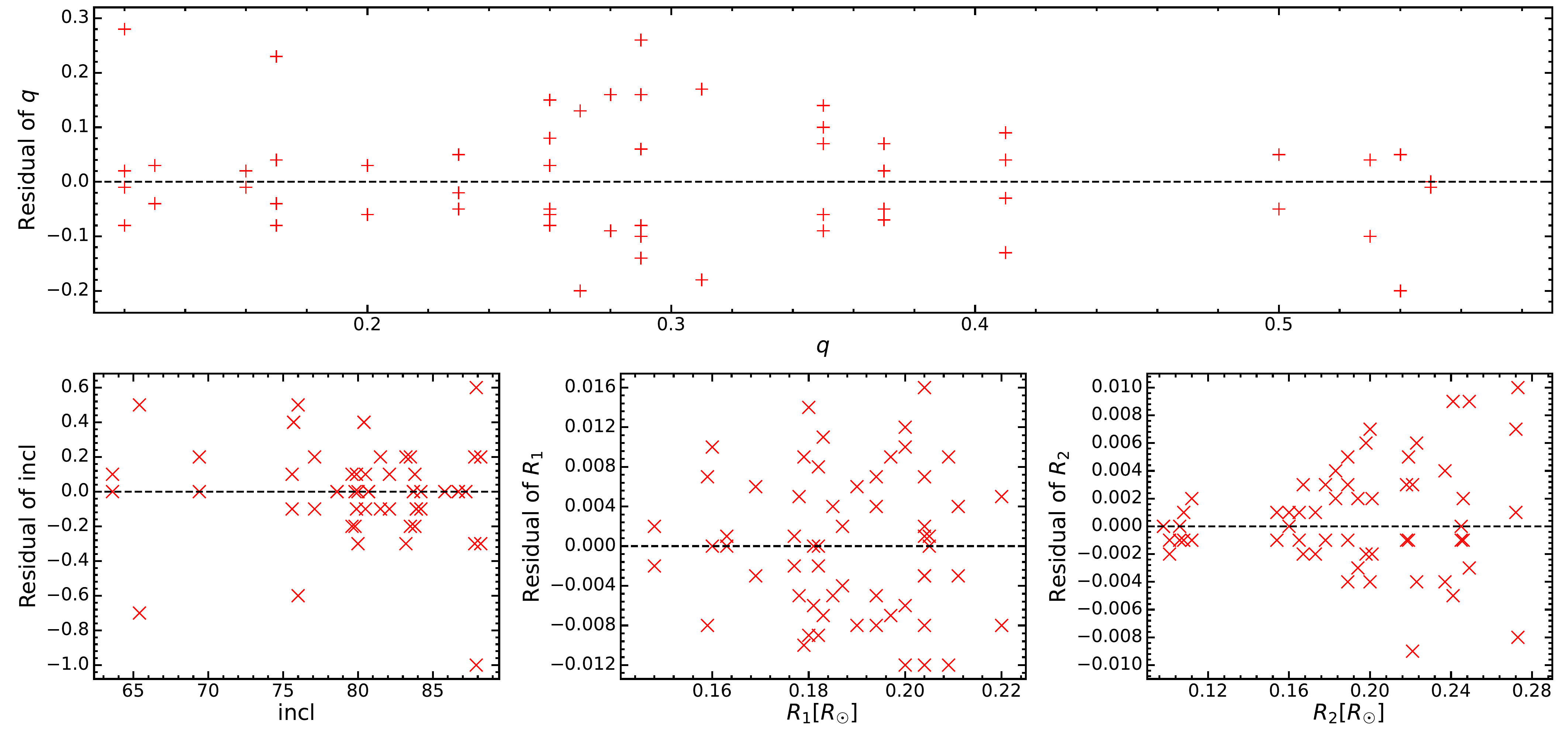}
\vspace{-0.0in}
\caption{\label{F6}The relationship between the mean values and the residuals when $q$ is fixed to the maximum error and minimum value, the top panel is for $q$. The left-bottom, middle-bottom, and the right-bottom panels are similar as the top but for $i$, $r_{1}$ and $r_{2}$, respectively.}
\end{figure*}

\begin{figure}
\begin{minipage}[t]{0.48\linewidth}

\centering
\vspace{-0.0in}
\includegraphics[angle=0,width=\linewidth]{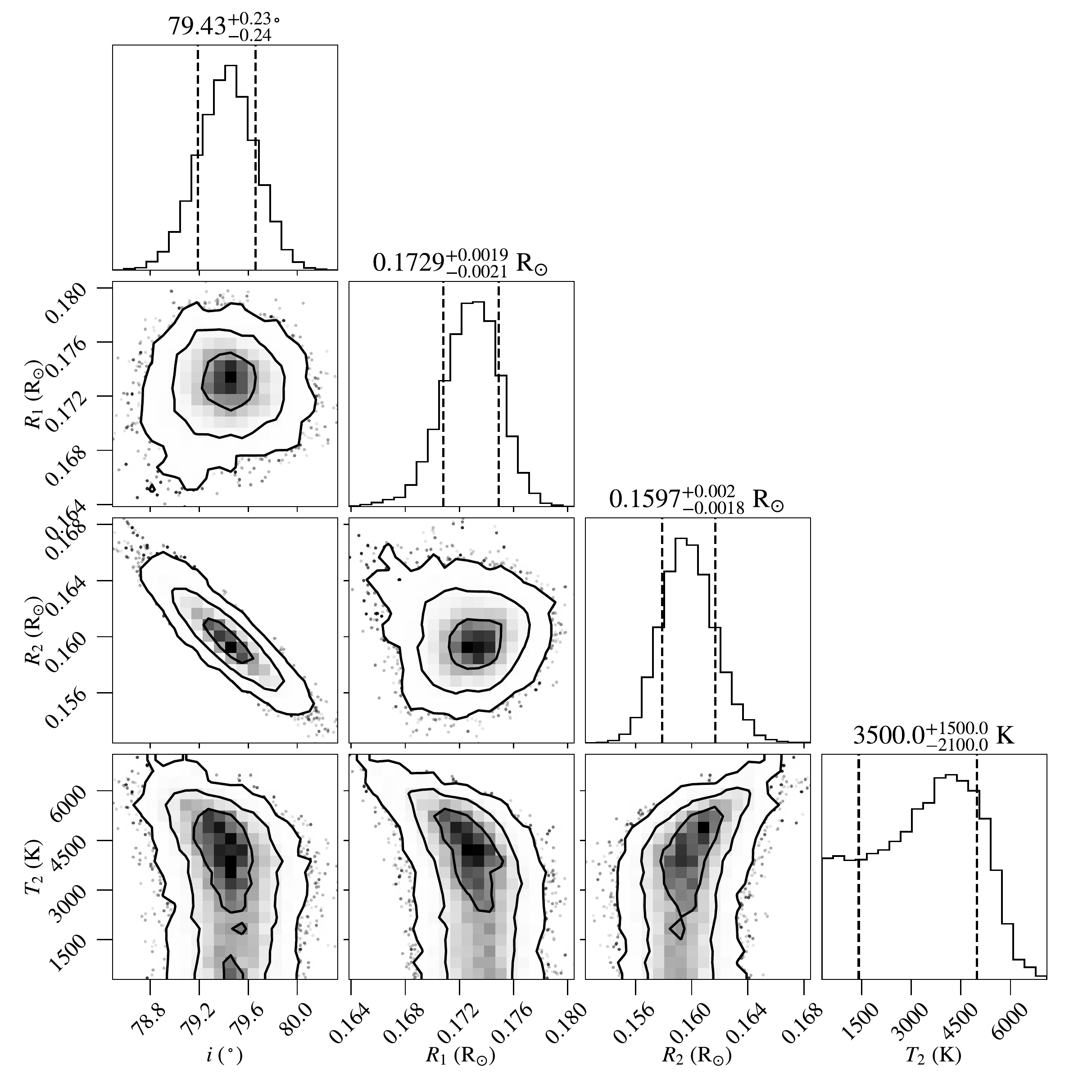}
\vspace{-0.0in}
\caption{\label{F7}MCMC results for ZTFJ011338.79+431154.9 when fixed $q$ is set to minimum error.}
\end{minipage}
\hspace{0.04\linewidth}
\begin{minipage}[t]{0.48\linewidth}

\centering
\vspace{-0.0in}
\includegraphics[angle=0,width=\linewidth]{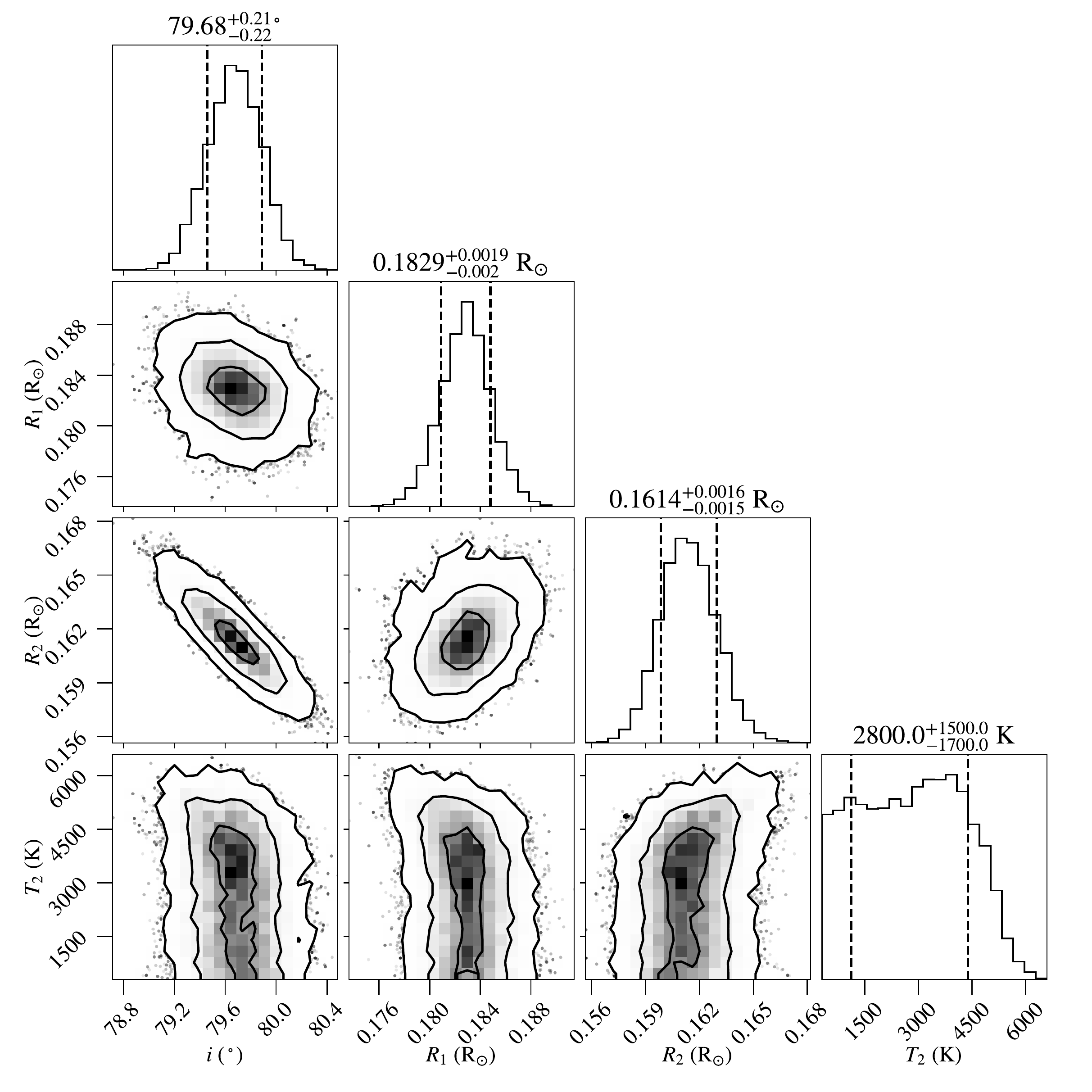}
\vspace{-0.0in}
\caption{\label{F8}Similar to Figure \ref{F7}, but fixed $q$ to maximum error.}
\end{minipage}
\end{figure}

Among the obtained orbital parameters, the mass ratio q has the largest error. To check whether the accuracy of the other parameters is affected by $q$, we fix $q$ to the maximum and minimum values ($q_{\max}=q+\sigma_q$, $q_{\min}=q-\sigma_q$ see section 4.1 and Table \ref{T2}) and performed the PHOEBE analysis again, respectively. Figure \ref{F7} and Figure \ref{F8} show the mcmc results when $q$ is fixed to the minimum and maximum values using ZTFJ011338.79+431154.9 as an example. The residuals of $i$, $r_{1}$ and $r_{2}$ were estimated by the difference between the new values and the mean values in Table \ref{T2}. In Figure \ref{F6}, the top panel shows that the relationship between the mean $q$ and the residual values of $q$, while the left-bottom, middle-bottom and right-bottom panels are similar, but for $i$, $r_{1}$ and $r_{2}$ respectively. The average percentage error of $q$ is 32.8 $\%$, but only 3.1 $\%$ and 1.4$\%$ for $r_{1}$ and $r_{2}$. These imply that for the sdB+dM system, the radii are reliable even if $q$ has a large error. We also noted that $i$, $r_{1}$, and $r_{2}$ all exhibit Gaussian-like distributions for a fixed $q$ in Figure \ref{F7} and Figure \ref{F8}.

The selection of initial values in PHOEBE might lead a bias to the solution of the orbital parameters. Our default initial values of $q$, $r_{1}$, and $r_{2}$ are 0.3, 0.175, and 0.15, where $q$ obeys a uniform distribution and $r_{1}$ and $r_{2}$ obey Gaussian distributions with a standard deviation of 0.05. To investigate the effect of the initial values on the final results, we also adopted an unreasonable initial values of $q$ = 0.1, $r_{1}$ = 0.05, and $r_{2}$ = 0.05. From Table \ref{T3}, we can see that the choice of the initial value has little effect on our results.

\begin{table}

\begin{center}
\caption[]{ZTF J184847.05+115720.3 as as an example to test the effect of different initial values.}\label{T3}

\begin{tabular}{ccc}
  \hline\noalign{\smallskip}
Initial values&Default&Unreasonable\\
  \hline\noalign{\smallskip}
$i$ ($^\circ$)&87.8$_{-0.6}^{+0.6}$&87.7$_{-0.5}^{+0.7}$ \\

$q$&0.35$_{-0.06}^{+0.07}$&0.32$_{-0.05}^{+0.09}$\\
$r_{1} ( R_{\odot})$&0.194$_{-0.006}^{+0.005}$&0.192$_{-0.005}^{+0.006}$\\
$r_{2} (R_{\odot})$&0.201$_{-0.002}^{+0.002}$&0.200$_{-0.002}^{+0.003}$\\
$T_{2} (K)$&2300$_{-1300}^{+1300}$&2300$_{-1400}^{+1400}$\\
  \hline\noalign{\smallskip}
\end{tabular}
\item {Note: $q$ obeys a uniform distribution while the others obey Gaussian distributions.}

\end{center}
\end{table}

\subsection{Comparison with previous work} %ZTF J162256.66+473051.1, ZTF J153349.44+375927.8 and ZTF J223421.49+245657.1
Three of our sdB+dM eclipsing binaries were studied in previous works, which used both the single-line radial velocity curve and LCs to determine parameters. We compared our results with theirs to validate our method. As shown in Table \ref{T4}, parameters of ZTF J162256.66+473051.1 (PG 1621+476) and ZTF J153349.44+375927.8 (FBS 1531+381) are well agreed with \citet{2014A&A...564A..98S} and \citet{2010ApJ...708..253F}, respectively. \citet{2010ApJ...708..253F} yielded a relatively small mass for the sdB star, but it does not conflict with the typical mass of sdB if the error is taken into account. We also testified that using a temperature of 30,000 K or temperatures from the literature hardly affects our parameter determination.

As shown in Table \ref{T5}, the case of ZTF J223421.49+245657.1 is more complicated. \citet{2007ASPC..372..483O} suggested that the primary component is an sdB star with a mass of 0.47 $M_{\odot}$ or 0.499 $M_{\odot}$, while \citet{2017MNRAS.472.3093A} suggested that it is a low-mass white dwarf with a mass of 0.19 $M_{\odot}$ or 0.288 $M_{\odot}$. Our $q$ agrees with these two works while $R_1, R_2$ do not. It is worth noting that the radii of \citet{2007ASPC..372..483O} were given in units of orbital separation not solar radius. Our radius ratio is consistent with \citet{2017MNRAS.472.3093A}, which means that the reason for the different parameters is due to the difference in the adoption of $M_1$. We prefer 
the primary component to be a typical sdB when considering its location on the CMD. 

In summary, when the primary star is a typical sdB, parameters $i$, $q$, $R_{1}$, $R_{2}$, and $M_{2}$ obtained from only LCs are consistent with those obtained from both LCs and the single-line radial velocity curve. With the single-line radial velocity curve, $q$ and $M_2$ can be determined with better precision, especially when the error of the mass ratio based on LCs alone is large. According to the discussion in Section 3 and 
this section, the mass estimates of sdB stars based on the single-line radial velocity curve have a large uncertainty if $q$ is not well constrained by LCs. 

\begin{table*}
\vspace{-0.0in}
\begin{center}

\caption[]{Physical parameters of ZTF J162256.66+473051.1 and ZTF J153349.44+375927.8 compared to previous works}\label{T4}

\vspace{0.0in}

\begin{tabular}{c|cc|cc}
\hline\noalign{\smallskip}
& \multicolumn{2}{c|}{ZTF J162256.66+473051.1 (PG 1621+476)} &\multicolumn{2}{c}{ZTF J153349.44+375927.8 (FBS 1531+381)}\\

\hline\noalign{\smallskip} 
$i$($^\circ$)&$69.4^{+0.5}_{-0.5}$&$72.33\pm{1.11}$&$86.7^{+0.1}_{-0.1}$&$86.6\pm{0.2}$\\
$q$&$0.13^{+0.04}_{-0.04}$&0.1325[fixed]&$0.37^{+0.02}_{-0.05}$&0.301\\

$R_1 $($R_{\odot}$)&$0.182^{+0.004}_{-0.004}$&$0.168\pm{0.007}$&$0.177^{+0.001}_{-0.002}$&$0.166\pm{0.007}$\\
$R_2$($R_{\odot}$)&$0.098^{+0.002}_{-0.002}$&$0.085\pm{0.004}$&$0.165^{+0.001}_{-0.002}$&$0.152\pm{0.005}$\\
$T_1$(K)&29000[fixed]&29000[fixed]&29230[fixed]&$29230$[fixed]\\
 
$T_2$(K)&$2700^{+1000}_{-1500}$&$2500\pm{900}$&$2700^{+800}_{-1400}$&$3100\pm{600}$\\
 
$M_{1}$($M_{\odot}$)&0.47[fixed]&$0.48\pm{0.03}$&0.47[fixed]&$0.376\pm{0.055}$\\
 
$M_{2}$($M_{\odot}$)&$0.061_{-0.019}^{+0.019}$ &$0.064\pm{0.04}$&$0.174_{-0.024}^{+0.009}$&$0.113\pm{0.017}$\\
 
$L_1$($L_{\odot}$)& $17^{+1}_{-1}$&...&$18.0^{+0.3}_{-1.0}$&$18.14\pm{1.84}$\\
$L_2$($L_{\odot}$)&$0.0003^{+0.0007}_{-0.001}$&...&$0.001^{+0.002}_{-0.003}$&...\\
   &this work& Schaffenroth et al. 2014 &this work& For et al. 2010\\
\hline\noalign{\smallskip}
\end{tabular}

\item[]{Note: parameters marked with [fixed] were fixed in MCMC estimations.}

\end{center}
\end{table*}

\begin{table*}
\vspace{-0.0in}
\begin{center}
\caption{\label{T5}Physical parameters of ZTF J223421.49+245657.1 compared to previous works}
\vspace{0.15in}

\begin{tabular}{c|ccccc}
\hline\noalign{\smallskip}
 & \multicolumn{5}{c}{ZTF J223421.49+245657.1 (HS 2231+2441)}\\
\hline\noalign{\smallskip}    
$i$($^\circ$)&$78.6_{-0.4}^{+0.4}$&79.6&79.1
&$79.4\pm{0.2}$&$79.6\pm{0.1}$\\
$q$&$0.2^{+0.03}_{-0.06}$&0.159&0.145
&0.190[fixed]&0.160[fixed]\\
$R_1 $($R_{\odot}$)
&$0.204^{+0.004}_{-0.006}$&0.250[s]&0.250[s]&
$0.144\pm{0.004}$&$0.165\pm{0.005}$\\
$R_2$($R_{\odot}$)&$0.108^{+0.003}_{-0.003}$&0.127[s]&0.129[s]&
$0.074\pm{0.004}$&$0.086\pm{0.004}$\\
$T_1$(K)&28500[fixed]&$28370\pm{80}$[fixed]&$28370\pm{80}$[fixed]&
28500[fixed]&28500[fixed]\\
$T_2$(K)&$3400^{+2100}_{-2100}$&...&...&
$3010\pm{460}$&$3410\pm{500}$\\
$M_{1}$($M_{\odot}$)&0.47[fixed]&0.470&0.499&
$0.190\pm{0.006}$&$0.288\pm{0.005}$\\
$M_{2}$($M_{\odot}$)&$0.094_{-0.028}^{+0.014}$&0.075&0.072&
$0.036\pm{0.004}$&$0.046\pm{0.004}$\\
$L_1$($L_{\odot}$)& $20.0^{+1.0}_{-2.0}$&...&...&
...&...\\
$L_2$($L_{\odot}$)&$0.001^{+0.003}_{-0.003}$ &...&...&
...&...\\
  &this work &Østensen et al. 2007&Østensen et al. 2007&
  Almeida et al. 2017&Almeida et al. 2017\\

\hline\noalign{\smallskip}
\end{tabular}

\item[]{Note: Parameters marked with [fixed] were fixed in MCMC estimations. Parameters marked with [s] are in the unit of orbital separation rather than solar radius.}

\end{center}
\end{table*}

\subsection{Evolutionary stage of sdB+dM}
\begin{figure}[h]
\begin{minipage}[t]{0.48\linewidth}
\centering
\vspace{-0.0in}
\includegraphics[angle=0,width=\linewidth]{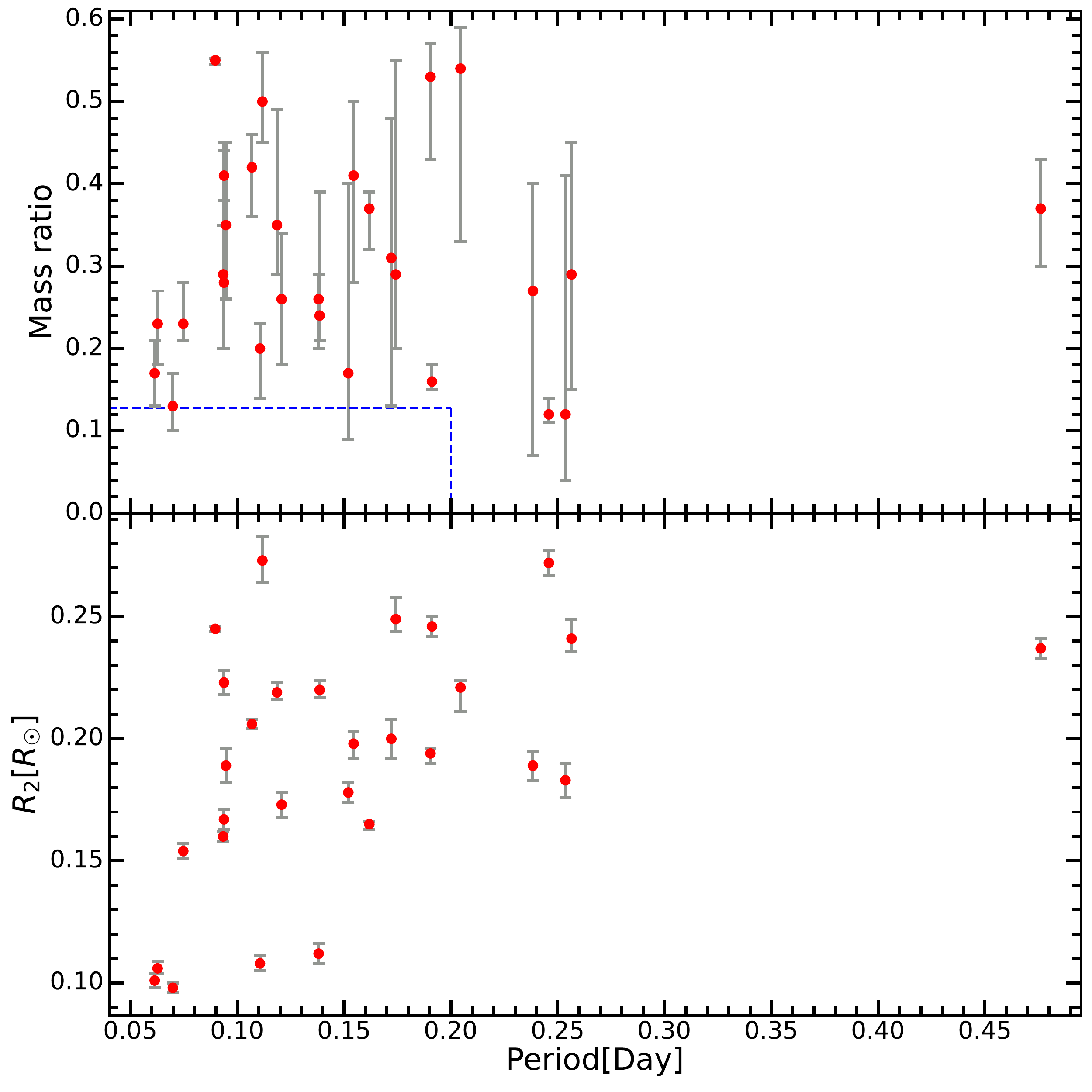}
\vspace{-0.0in}
\caption{\label{F9}\small Evolutionary diagram of 29 sdB+dM. The distribution of mass ratio with period is in the top panel, while the distribution of the companion's radius $R_{2}$ with period in the bottom panel. Our sdB+dM are shown as red dots. The blue dashed lines denote the period and mass limits of companion stars.}
  \end{minipage}%
  \hspace{0.04\linewidth}
  \begin{minipage}[t]{0.48\textwidth}
\centering
\vspace{-0.0in}
\includegraphics[angle=0,width=\linewidth]{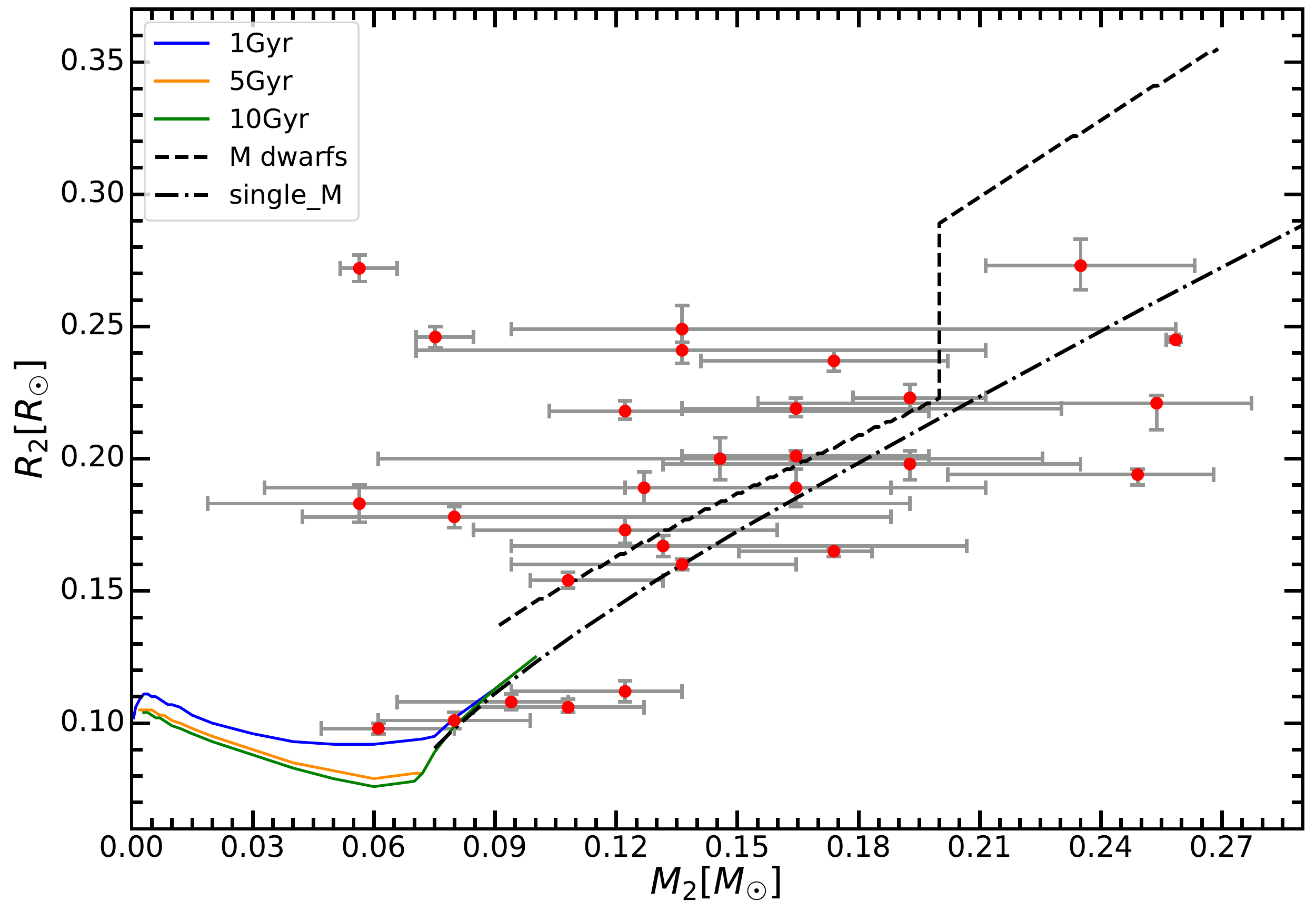}
\vspace{-0.0in}
\caption{\label{F10}\small Diagram of the mass--radius relation of the companion stars. The blue, orange and green solid lines denote theoretical mass--radius relations for brown dwarfs with ages of 1 Gyr, 5 Gyr and 10 Gyr. The black dotted line is theoretical mass--radius relation for M dwarfs in CVs. The black dot–dashed line represents the theoretical  mass-radius relation for single M-type stars. Our 29 sdB+dM are shown as red dots.}
  \end{minipage}%
\end{figure}

Figure \ref{F9} shows the evolutionary diagram of our sdB+dM eclipsing binaries. In the top panel of Figure \ref{F9}, the mass ratios are randomly distributed in $q=0.1-0.6$ when the orbital periods are longer than 0.1 days. However, we only found low mass-ratio ($q\sim0.2$) sdB+dM eclipsing binaries when $P<0.08$ days. In the bottom panel of Figure \ref{F6}, the decrease of $R_2$ with the orbital period is more significant than the decrease of $q$, given the much smaller uncertainty of $R_2$. According to Figure \ref{F6}, we found that in sdB+dM, the companion stars are found to be less massive as the orbital period shortens. This implies that sdB+dM with more massive companions merge earlier in the orbital decay process. When $P<0.075$ days, the companions of all three sdB+dM in this work are brown dwarfs. \citet{2015AN....336..437G} suggested that the orbital period $P< \sim 0.2$ d and the minimum mass of companion stars $M_{2}< \sim 0.06$ $M\odot$ is a critical region where no companion stars would survive in the common envelope phase. Instead they may merge with the sdB, or be evaporated \citep{1998AJ....116.1308S}. In the top panel of Figure \ref{F9}, the blue dashed lines are the period and mass limits of the companion stars, and the absence of stars in this region suggests that our results support Geier's arguments. More sdB+dM eclipsing binaries with periods shorter than 1 hour will be detected in future data releases of ZTF, and we will gain more knowledge about the evolutionary end of sdB+dM.

\subsection{Physical properties of companion stars}
Figure \ref{F10} shows the mass--radius relation of the companion stars. The blue, yellow and green solid lines are the theoretical mass--radius relation for brown dwarfs with ages of 1 Gyr, 5 Gyr and 10 Gyr \citep{2003A&A...402..701B}. The black solid line is mass--radius relation for M dwarfs in CVs from \citet{2011ApJS..194...28K}, The black dot–dashed line represents the theoretical  mass-radius relation for single M-type stars from \citet{1998A&A...337..403B}. Most of companion stars in our sample agree with the mass--radius relations. The three exceptions are due to the underestimation of the mass ratio error or not a genuine sdB+dM that host a 0.47 $M_\odot$ sdB. A gap was found between $R_2=0.12R_\odot$ and $R_2=0.15R_\odot$ and this gap is likely the boundary of low-mass stars and brown dwarfs. Our comparison reveals that the dM mass distribution in our sample is similar to that of \citet{2015A&A...576A..44K}. Besides, the absolute magnitude and intrinsic color of five companion stars with $R_{2} < \sim 0.12 R_{\odot}$ are uncharacteristic. These suggest that the gap is not the result of a selection effect. In the future, based on hundreds of sdB+dM eclipsing binaries, it will be possible to update the mass--radius relation for low-mass stars and brown dwarfs. This will help to study the minimum mass of low-mass stars, and the maximum mass of brown dwarfs.

\section{Conclusion}

In this work, we have selected a sample of 33 HW Vir-type stars in ZTF DR5. Based on {\sl Gaia} EDR3 parallax and extinction correction, we found that these HW Vir-type stars are concentrated in a clump in the intrinsic color vs. absolute magnitude diagram. Their locations in the CMD imply that they are sdB+dM eclipsing binaries. By fixing the mass and temperature of sdB to $M_1=0.47M_\odot$ and $T_1=30,000$ K and setting the prior of $q$ and $T_2$, we analyzed LCs using PHOEBE 2.3 code to obtain the probability distributions of parameters $q$, $R_1$, $R_2$, $i$, and $T_2$. We obtained LC solutions for 29 sdB+dM eclipsing binaries with full primary eclipse detection. $R_1$, $R_2$, and $i$ obey a Gaussian-like distribution and have little correlation with other parameters, which means that they are well constrained by the LC analysis. $q$ does not show a Gaussian distribution in most cases, and the mean uncertainty is 0.08. 

Our parameters for three sdB+dM are comparable with previous works that used both LCs and the single-line radial velocity curve if the mass of sdB is $M_1=0.47M_\odot$. This means that parameters of sdB+dM determined from LCs are suitable for statistical analysis. Based on 29 sdB+dM, we found that both $q$ and $R_2$ decrease with the decreasing of the orbital period. sdB+dM with larger mass companions are likely to merge early during the shortening of the orbit. It is worth mentioning that companions of all three sdB+dM are brown dwarfs when the orbital period is less than 0.075 days. The masses and radii of the companion stars are consistent with the mass--radius relation for low-mass stars and brown dwarfs. We found a gap between $R=0.12R_\odot$ and $R=0.15R_\odot$ which can be explained as the boundary between low-mass stars and brown dwarfs.

sdB+dM eclipsing binaries are important objects to study the nature of sdB and dM, and their evolutionary endings are very interesting. With deeper photometry and more detections, ZTF will detect and classify hundreds of short-period sdB+dM eclipsing binaries. Before the availability of large-scale and deep spectroscopic surveys, the statistical properties of sdB and dM can be obtained from the LC analysis of large samples.

\begin{acknowledgements}
We thank the anonymous referee for the useful comments.
This work is supported by Sichuan Science and Technology Program (Grant No. 2020YFSY0034) and National Natural Science Foundation of China (NSFC) through the projects 12003022, 12173047, 11903045, 12003046, and U1731111. This work is also supported by Major Science and Technology Project of Qinghai Province 2019-ZJ-A10. 

This work has made use of PHOEBE software for the analysis of the light curve. PHOEBE is funded in part by the National Science Foundation (NSF \#1517474, \#1909109) and the National Aeronautics and Space Administration (NASA 17-ADAP17-68). The PHOEBE project web site is http://phoebe-project.org/. This work has made use of data from the European Space Agency (ESA) mission
{\it Gaia} (\url{https://www.cosmos.esa.int/gaia}), processed by the {\it Gaia}
Data Processing and Analysis Consortium (DPAC,
\url{https://www.cosmos.esa.int/web/gaia/dpac/consortium}). Funding for the DPAC
has been provided by national institutions, in particular the institutions
participating in the {\it Gaia} Multilateral Agreement.
This publication is based on observations
obtained with the Samuel Oschin 48-inch Telescope at the
Palomar Observatory as part of the Zwicky Transient Facility
project. ZTF is supported by the National Science Foundation
under grant AST-1440341 and a collaboration including Caltech,
IPAC, the Weizmann Institute for Science, the Oskar Klein Center
at Stockholm University, the University of Maryland, the
University of Washington, Deutsches Elektronen-Synchrotron
and Humboldt University, Los Alamos National Laboratories, the
TANGO Consortium of Taiwan, the University of Wisconsin at
Milwaukee, and Lawrence Berkeley National Laboratories.
Operations are conducted by COO, IPAC, and UW.

\end{acknowledgements}

\bibliographystyle{raa}
\bibliography{bibtex}

\appendix                  %%appendicial material is supported

\section{Figures for sdB+dM with Gaussian-like distributed mass ratio.}

\begin{figure}
\centering
\vspace{-0.0in}
\includegraphics[angle=0,width=85mm]{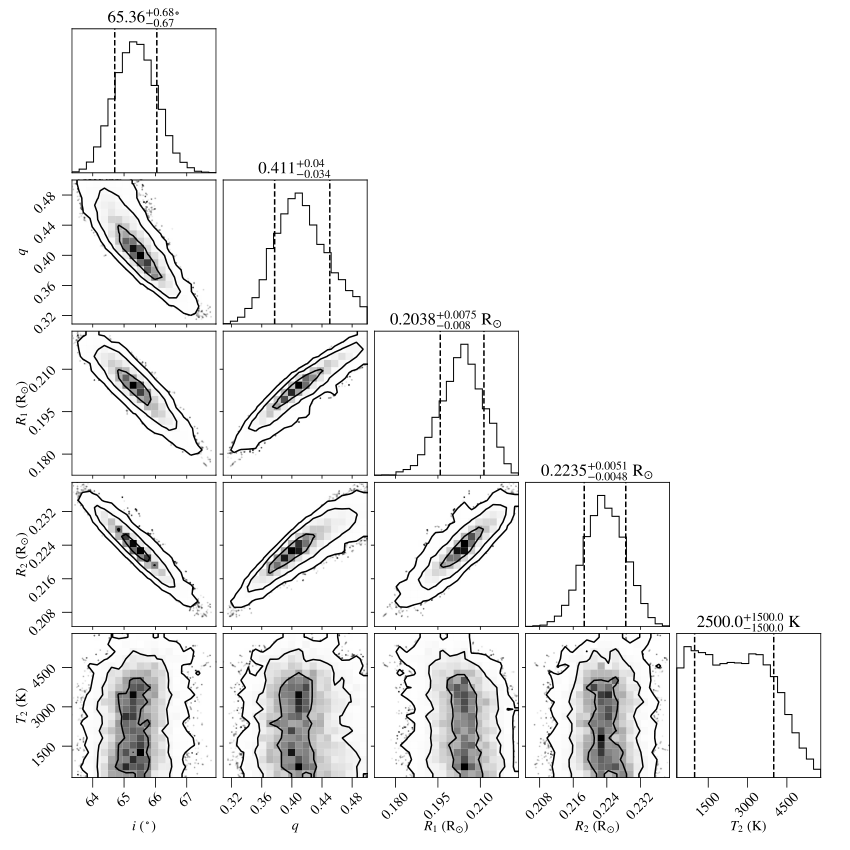}
\includegraphics[angle=0,width=64.06mm,height=85mm]{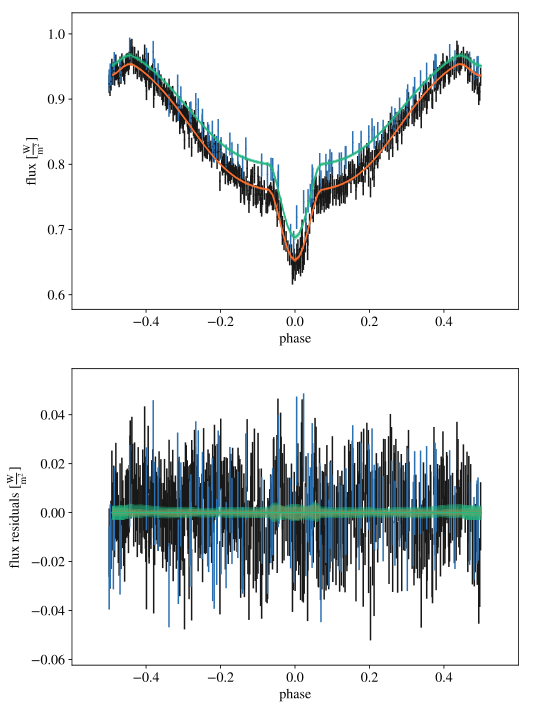}
\vspace{-0.0in}
\caption{\label{}Diagram of MCMC results and LC fitting for ZTF J072905.44-183703.4. }
\end{figure}

%%%%

\begin{figure}
\centering
\vspace{-0.0in}
\includegraphics[angle=0,width=85mm]{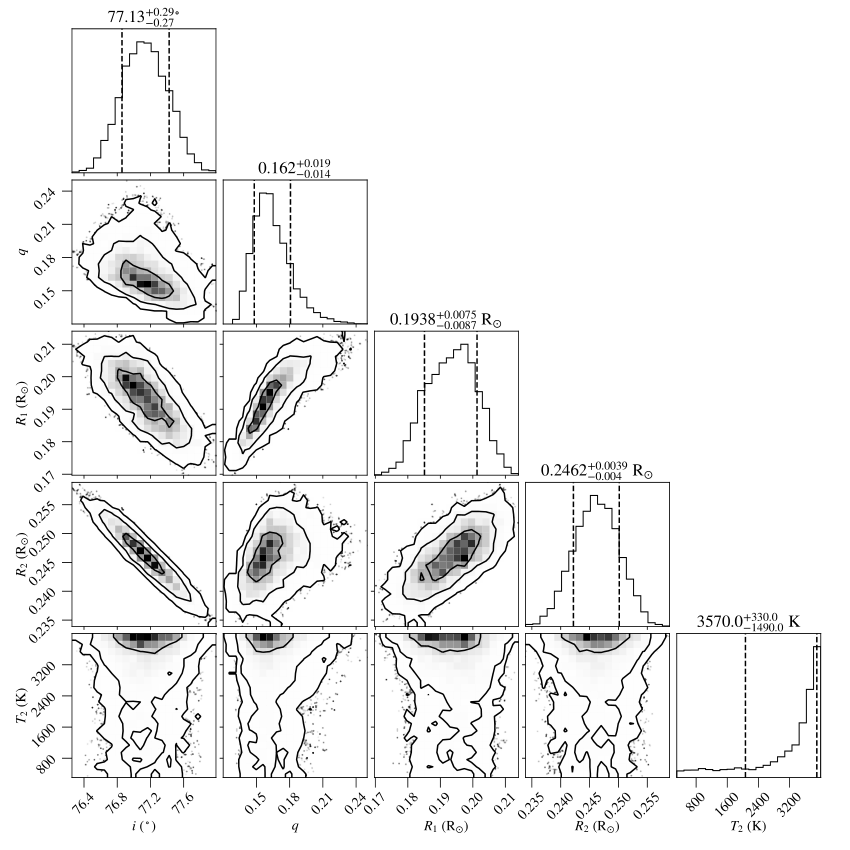}
\includegraphics[angle=0,width=64.06mm,height=85mm]{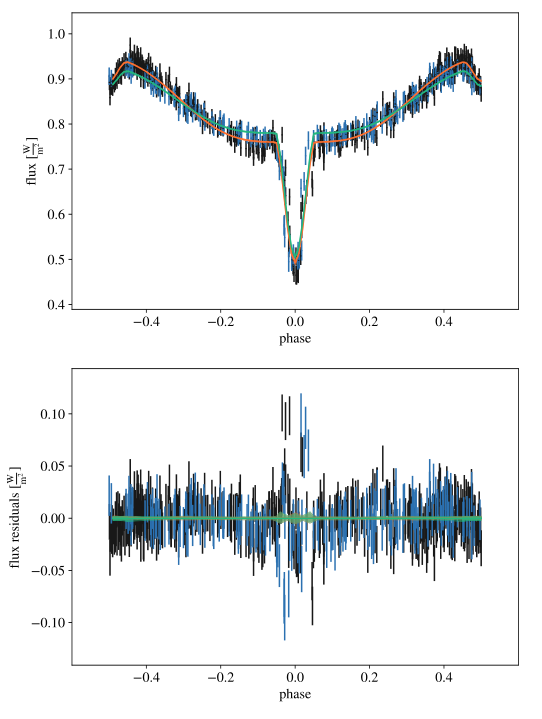}
\vspace{-0.0in}
\caption{\label{}Diagram of MCMC results and LC fitting for ZTF J184042.41+070321.9.}
\end{figure}

\begin{figure}
\centering
\vspace{-0.0in}
\includegraphics[angle=0,width=85mm]{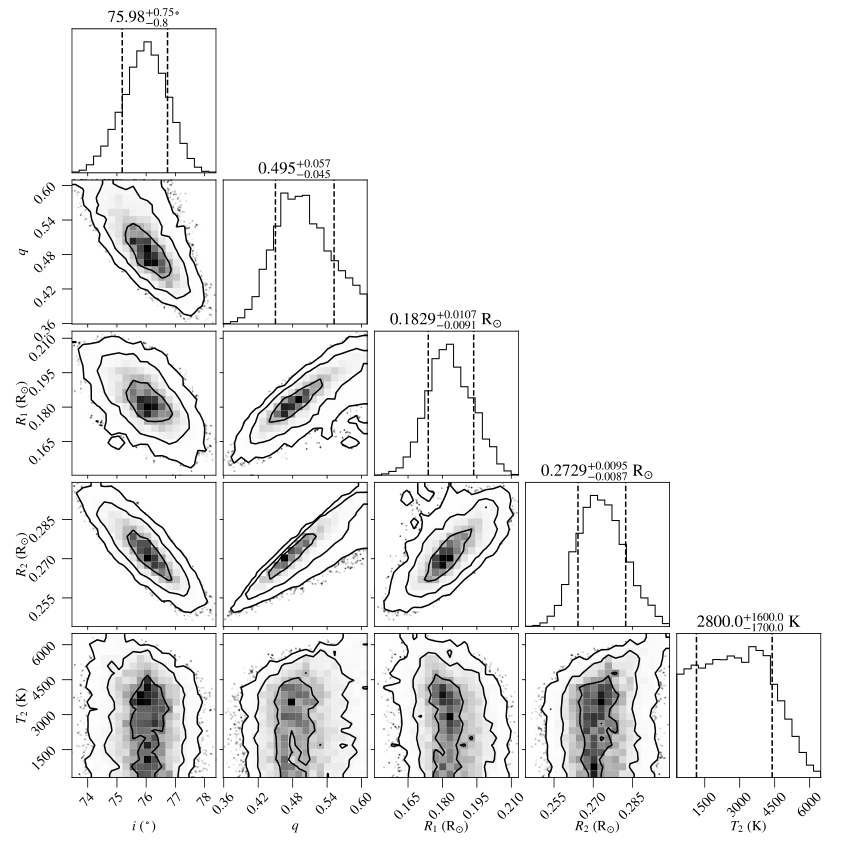}
\includegraphics[angle=0,width=64.06mm,height=85mm]{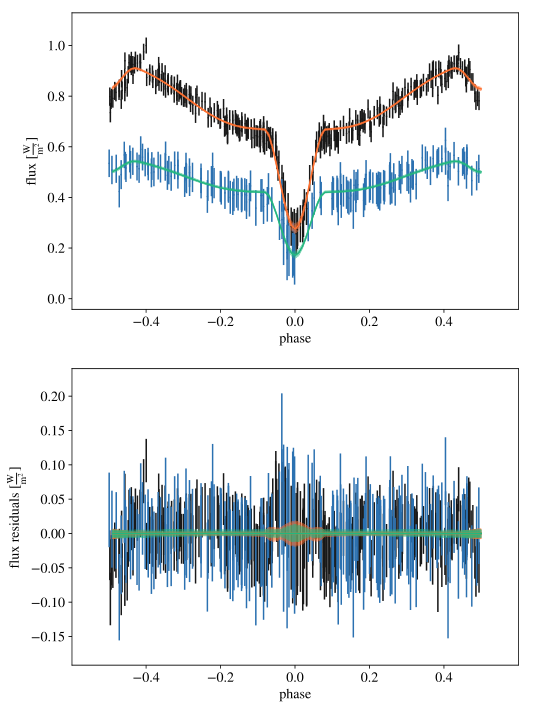}
\vspace{-0.0in}
\caption{\label{}Diagram of MCMC results and LC fitting for ZTF J192055.46+041619.5.}
\end{figure}

\section{Figures for sdB+dM with non-Gaussian distributed mass ratio.}

\begin{figure}
\centering
\vspace{-0.0in}
\includegraphics[angle=0,width=85mm]{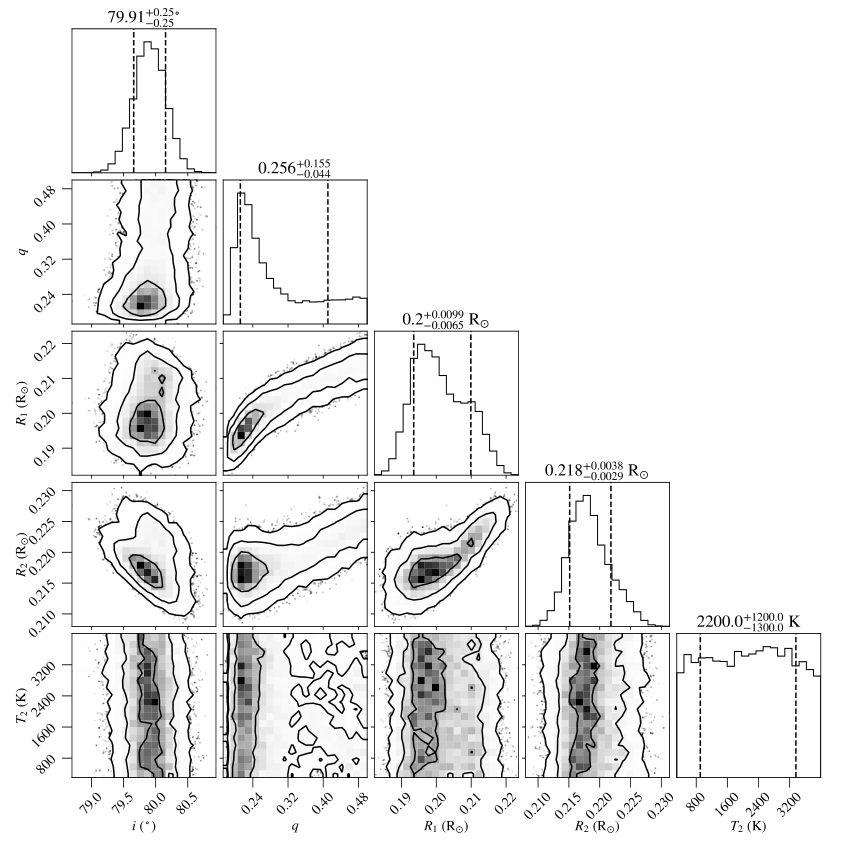}
\includegraphics[angle=0,width=64.06mm,height=85mm]{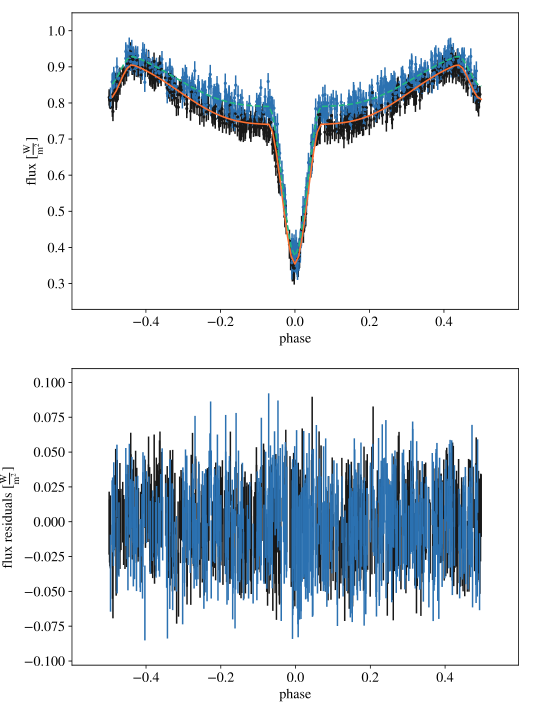}
\vspace{-0.0in}
\caption{\label{}Diagram of MCMC results and LC fitting for ZTF J050114.39+424741.3.}
\end{figure}

\begin{figure}
\centering
\vspace{-0.0in}
\includegraphics[angle=0,width=85mm]{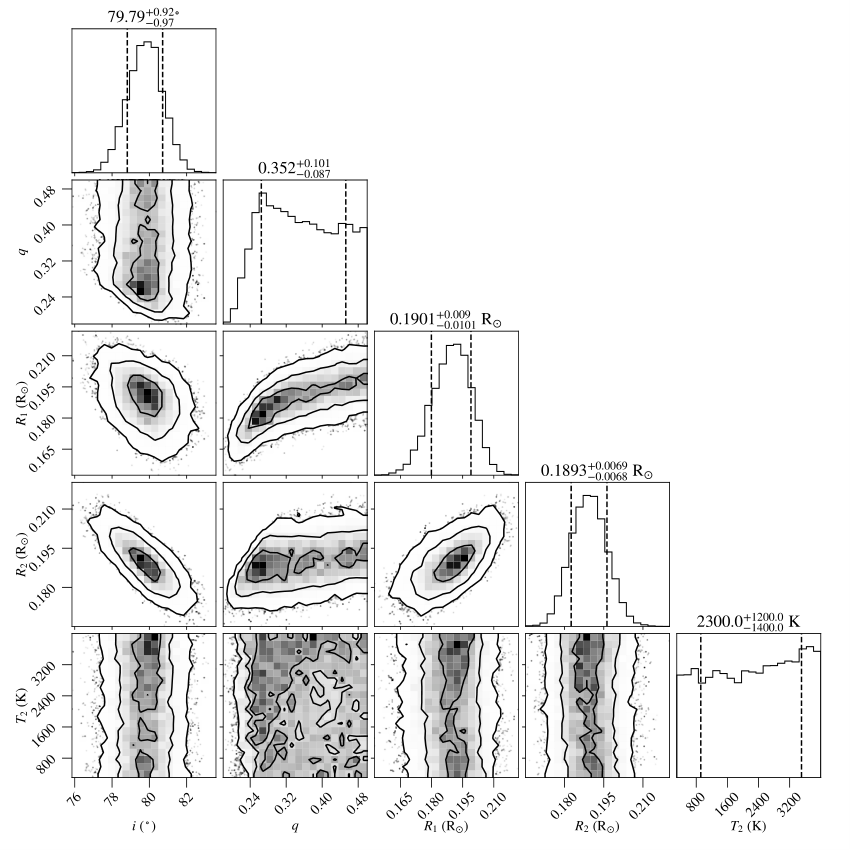}
\includegraphics[angle=0,width=64.06mm,height=85mm]{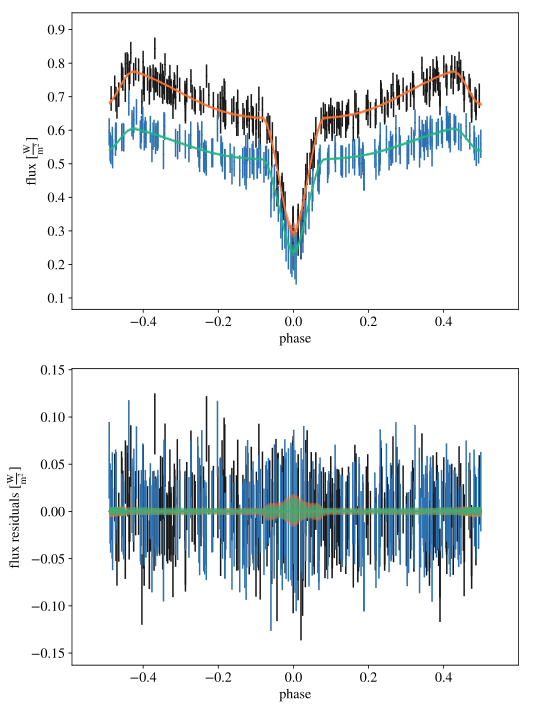}
\vspace{-0.0in}
\caption{\label{}Diagram of MCMC results and LC fitting for ZTF J054744.02+304732.2.}
\end{figure}

\begin{figure}
\centering
\vspace{-0.0in}
\includegraphics[angle=0,width=85mm]{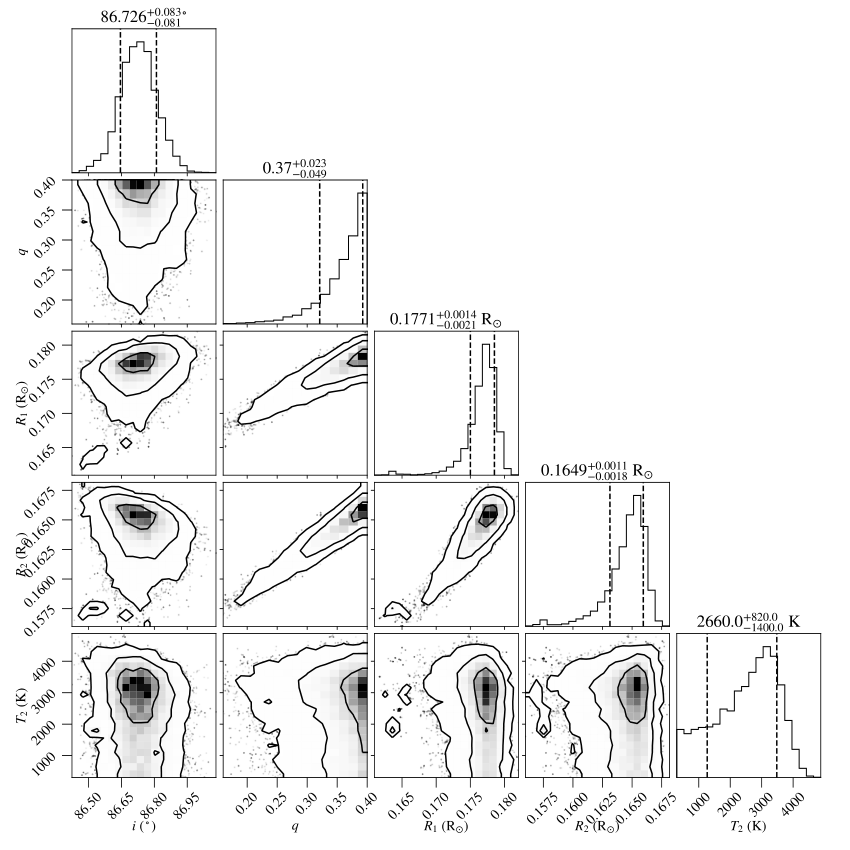}
\includegraphics[angle=0,width=64.06mm,height=85mm]{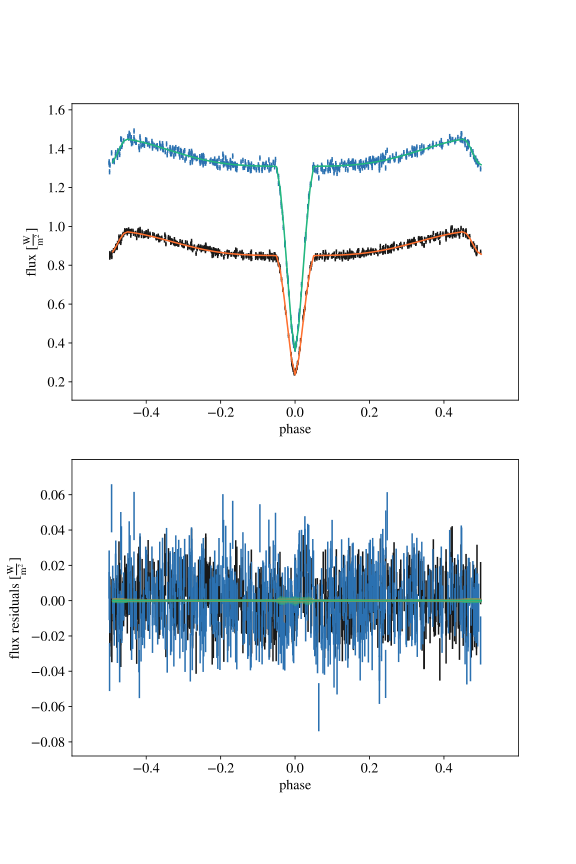}
\vspace{-0.0in}
\caption{\label{}Diagram of MCMC results and LC fitting for ZTF J153349.44+375927.8.}
\end{figure}

\begin{figure}
\centering
\vspace{-0.0in}
\includegraphics[angle=0,width=85mm]{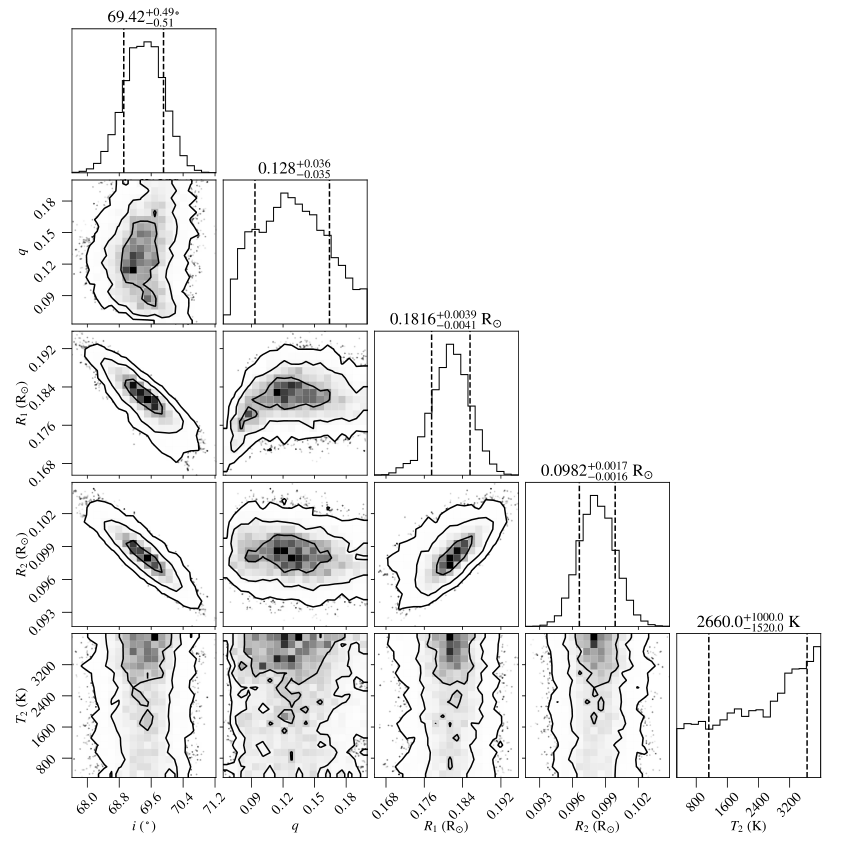}
\includegraphics[angle=0,width=64.06mm,height=85mm]{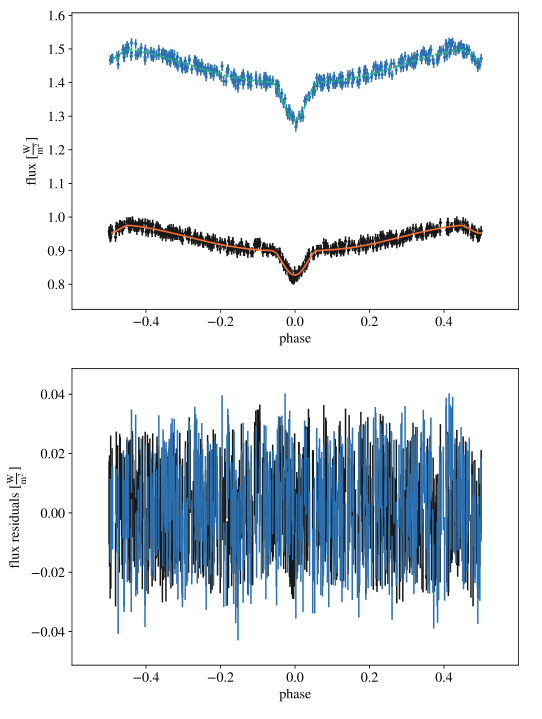}
\vspace{-0.0in}
\caption{\label{}Diagram of MCMC results and LC fitting for ZTF J162256.66+473051.1.}
\end{figure}

\begin{figure}
\centering
\vspace{-0.0in}
\includegraphics[angle=0,width=85mm]{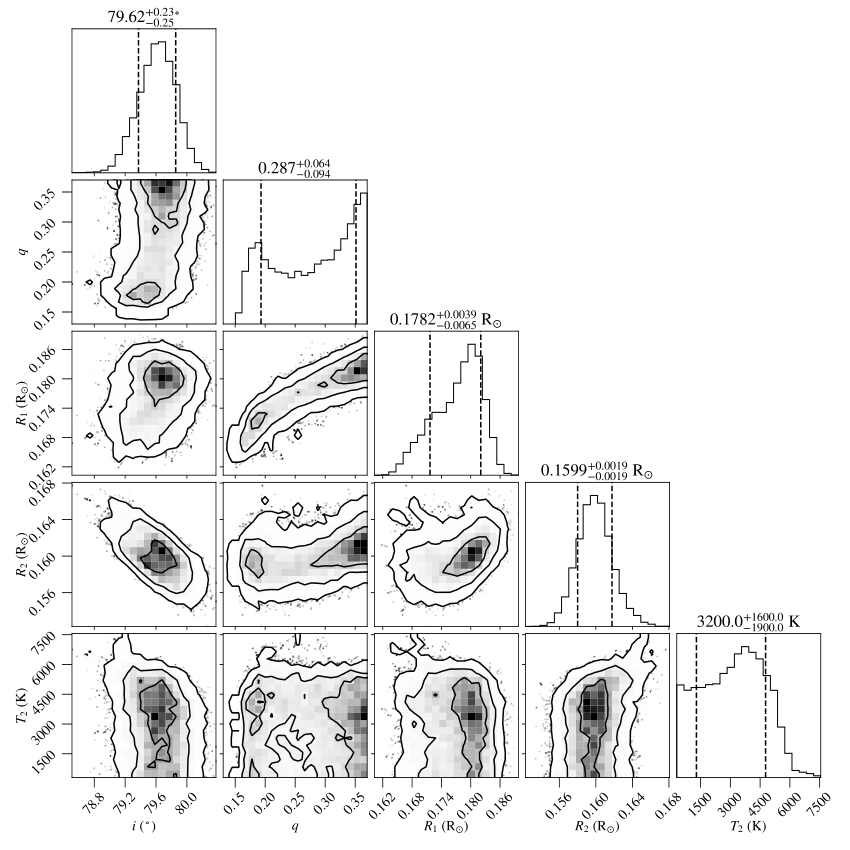}
\includegraphics[angle=0,width=64.06mm,height=85mm]{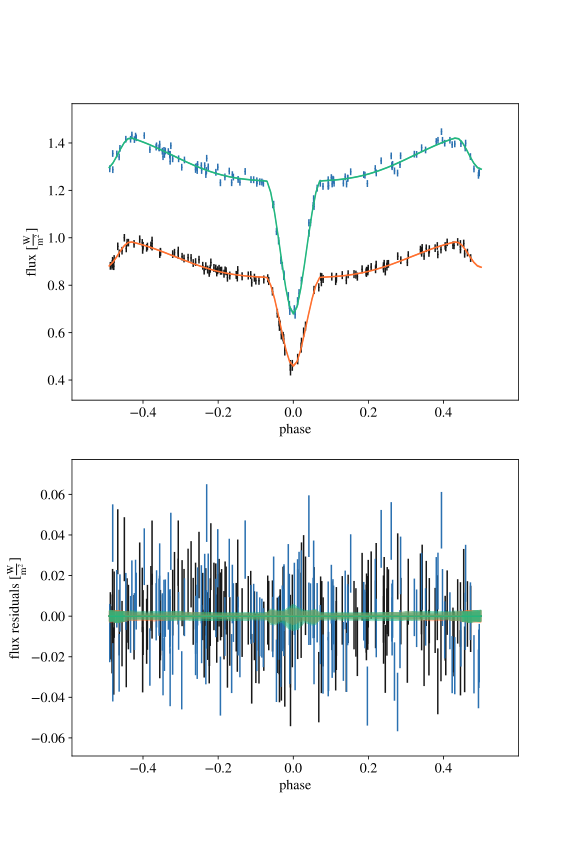}
\vspace{-0.0in}
\caption{\label{}Diagram of MCMC results and LC fitting for ZTF J011339.09+225739.0.}
\end{figure}

\begin{figure}
\centering
\vspace{-0.0in}
\includegraphics[angle=0,width=85mm]{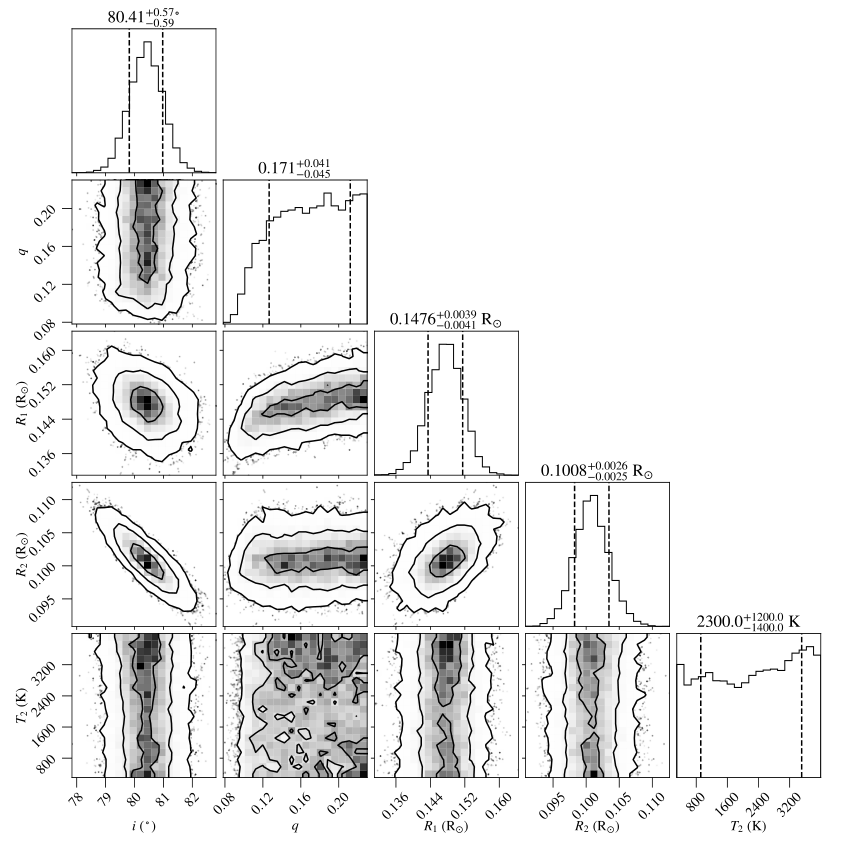}
\includegraphics[angle=0,width=64.06mm,height=85mm]{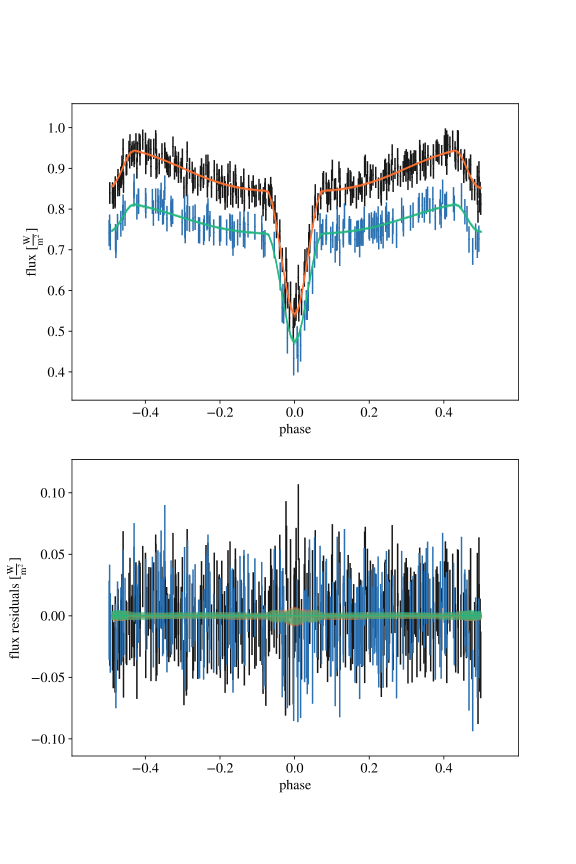}
\vspace{-0.0in}
\caption{\label{}Diagram of MCMC results and LC fitting for ZTF J183431.88+061056.7.}
\end{figure}

\begin{figure}
\centering
\vspace{-0.0in}
\includegraphics[angle=0,width=85mm]{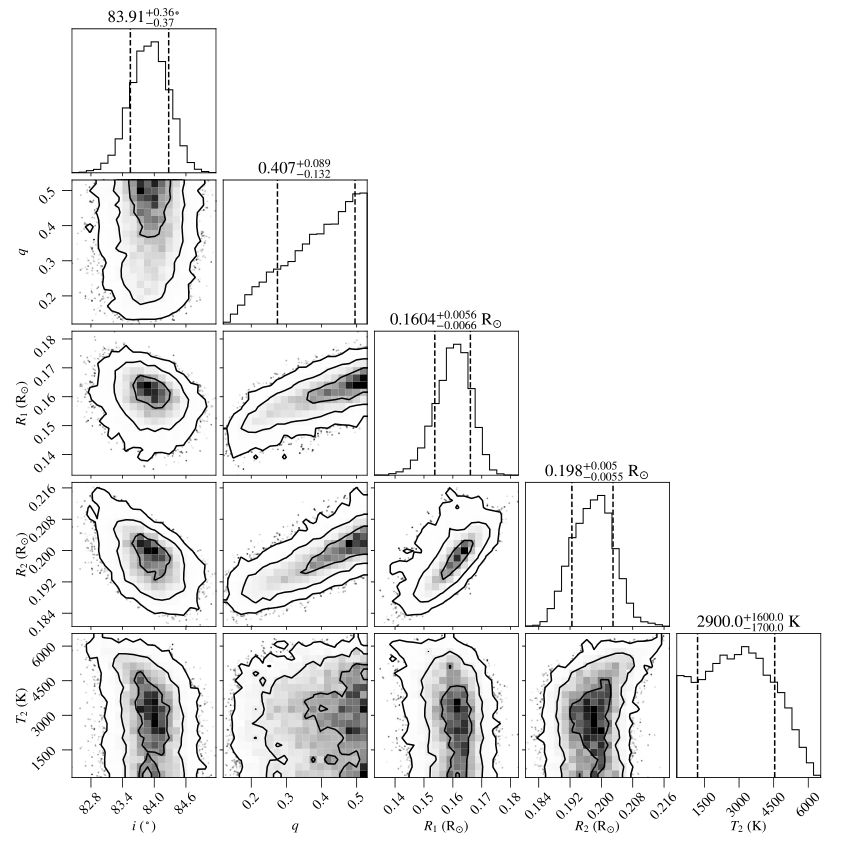}
\includegraphics[angle=0,width=64.06mm,height=85mm]{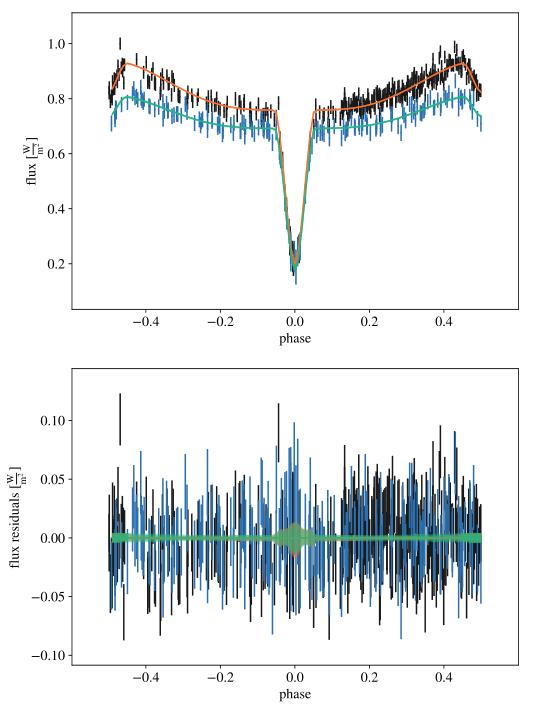}
\vspace{-0.0in}
\caption{\label{}Diagram of MCMC results and LC fitting for ZTF J183522.73+064247.1.}
\end{figure}

\begin{figure}
\centering
\vspace{-0.0in}
\includegraphics[angle=0,width=85mm]{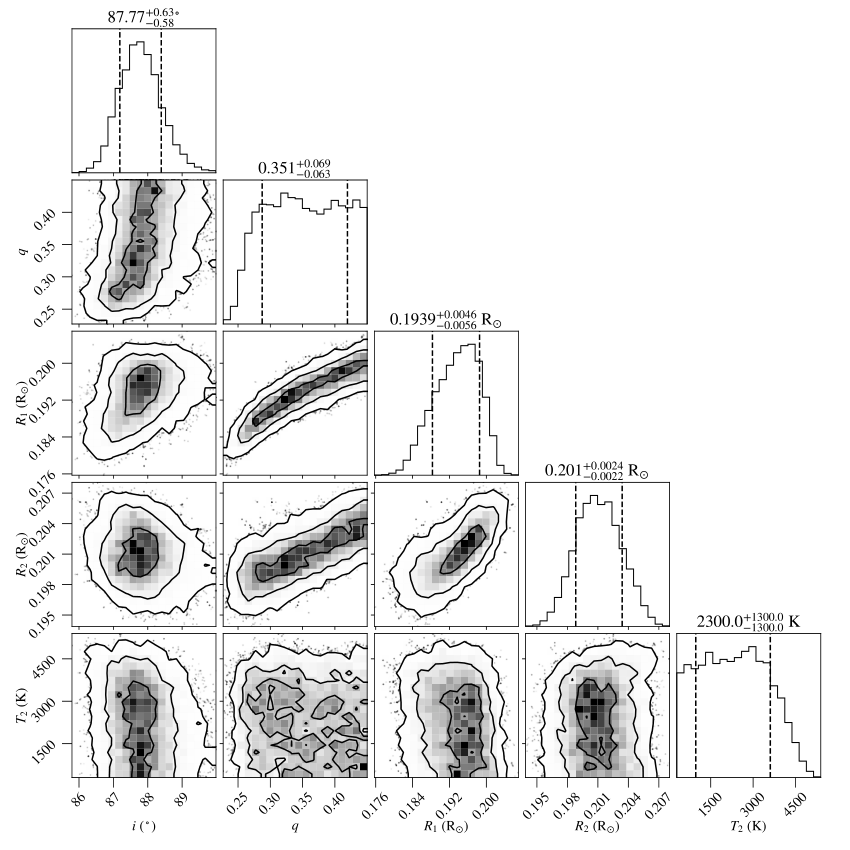}
\includegraphics[angle=0,width=64.06mm,height=85mm]{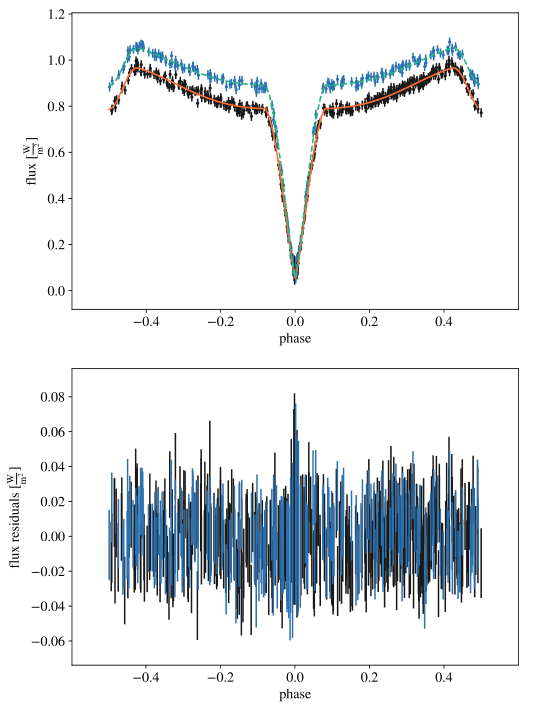}
\vspace{-0.0in}
\caption{\label{}Diagram of MCMC results and LC fitting for ZTF J184847.05+115720.3.}
\end{figure}

\begin{figure}
\centering
\vspace{-0.0in}
\includegraphics[angle=0,width=85mm]{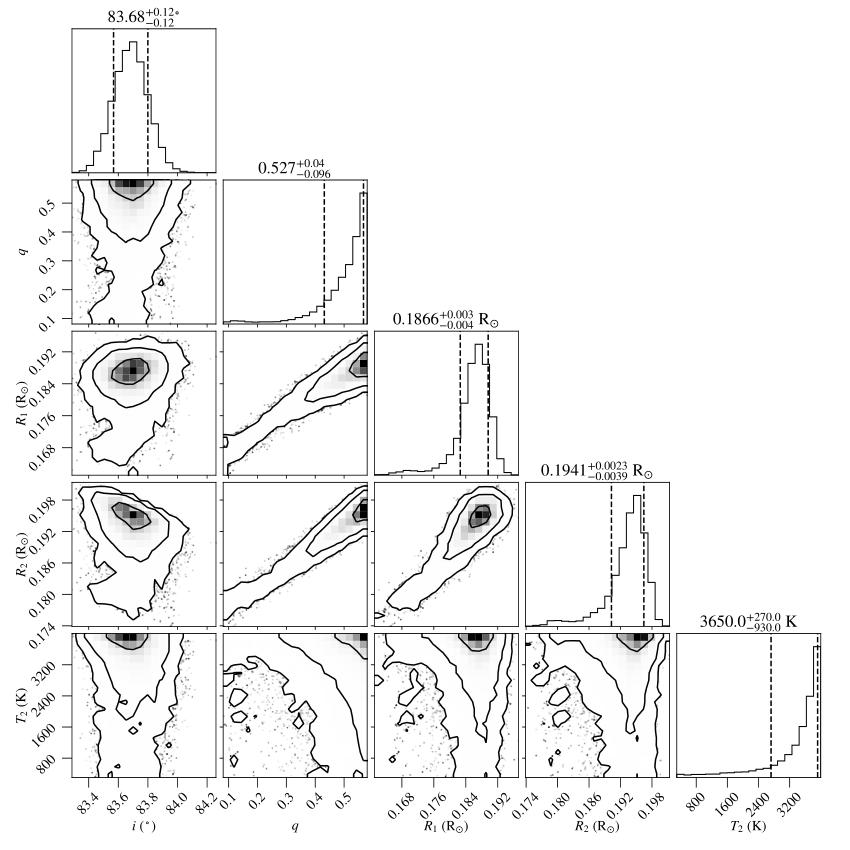}
\includegraphics[angle=0,width=64.06mm,height=85mm]{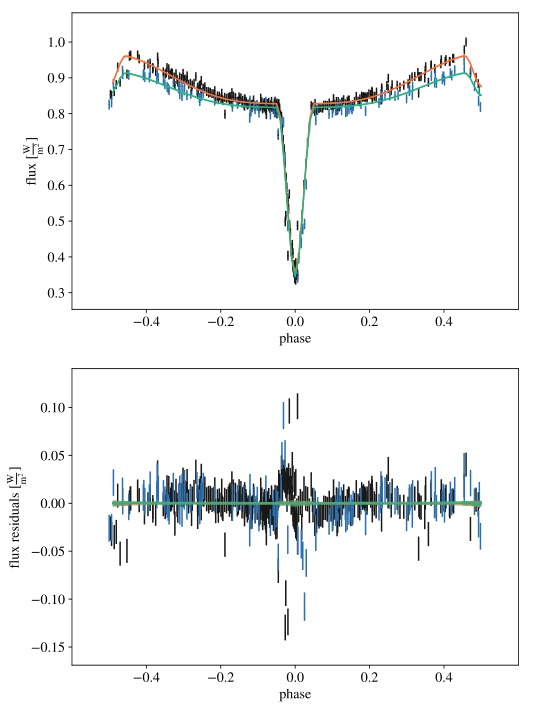}
\vspace{-0.0in}
\caption{\label{}Diagram of MCMC results and LC fitting for ZTF J185207.60+144547.1.}
\end{figure}

\begin{figure}
\centering
\vspace{-0.0in}
\includegraphics[angle=0,width=85mm]{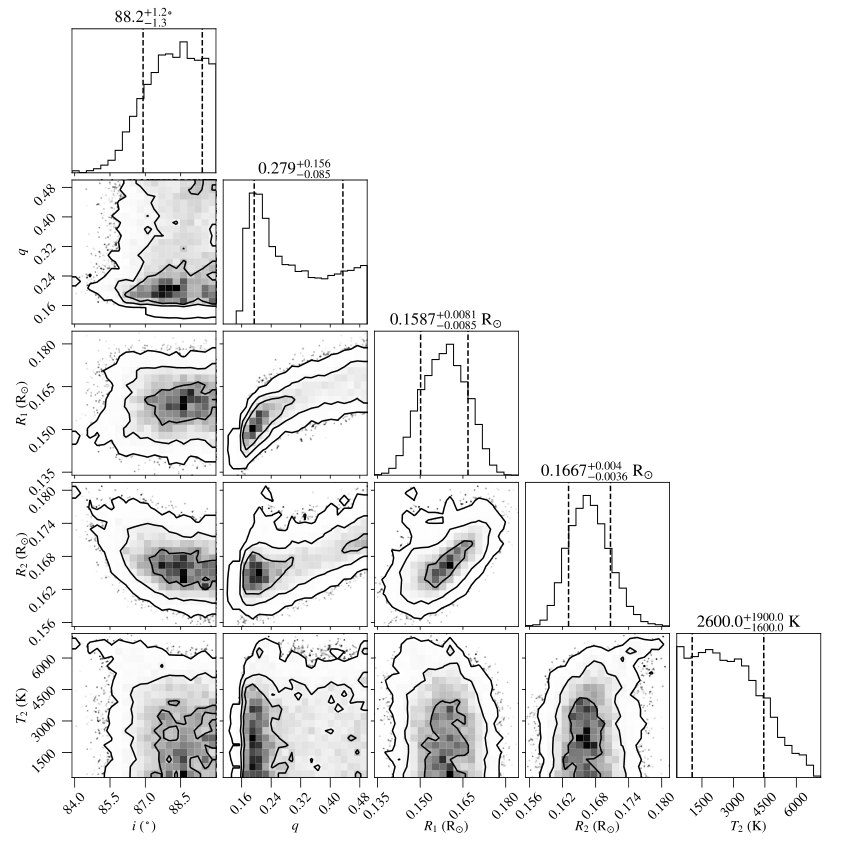}
\includegraphics[angle=0,width=64.06mm,height=85mm]{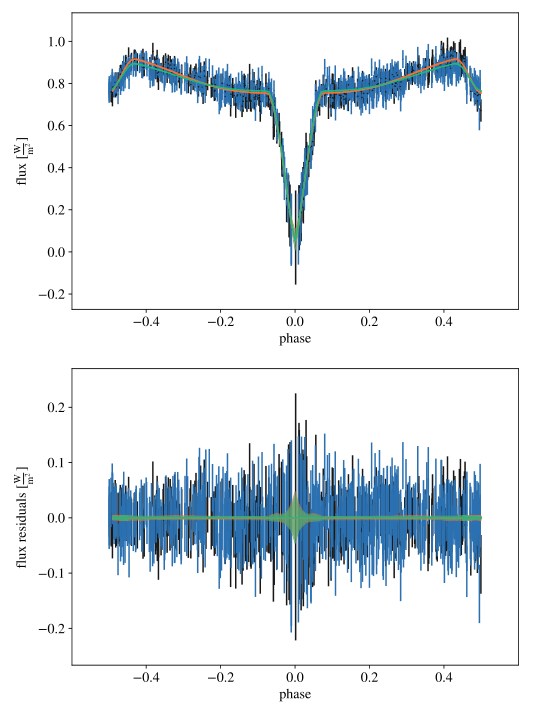}
\vspace{-0.0in}
\caption{\label{}Diagram of MCMC results and LC fitting for ZTF J014600.90+581420.4.}
\end{figure}
\begin{figure}
\centering
\vspace{-0.0in}
\includegraphics[angle=0,width=85mm]{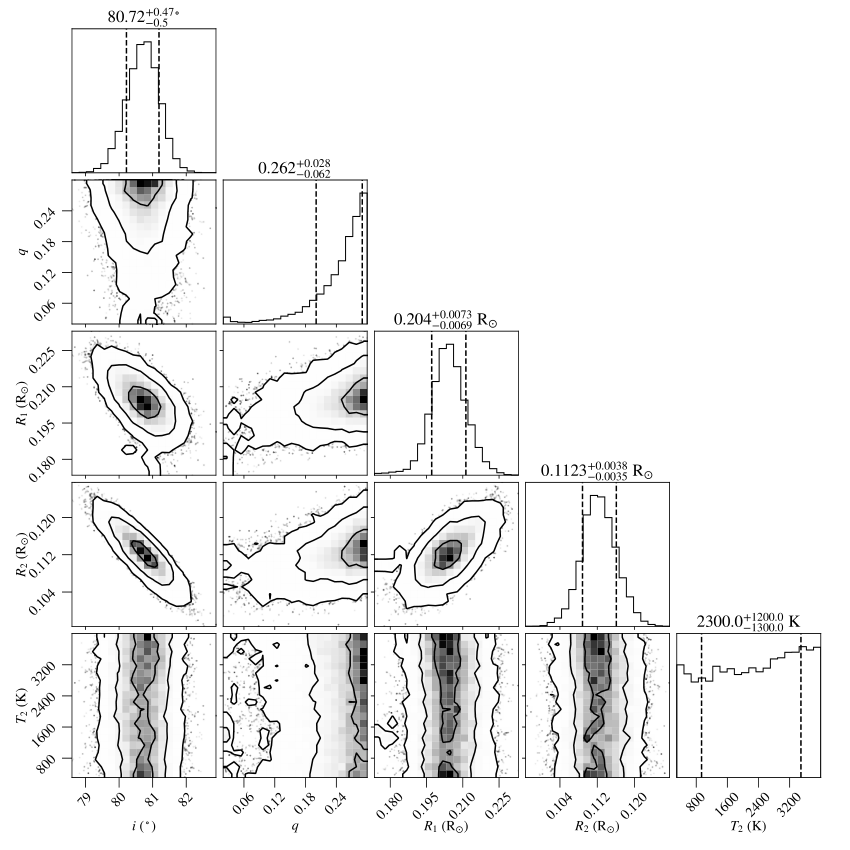}
\includegraphics[angle=0,width=64.06mm,height=85mm]{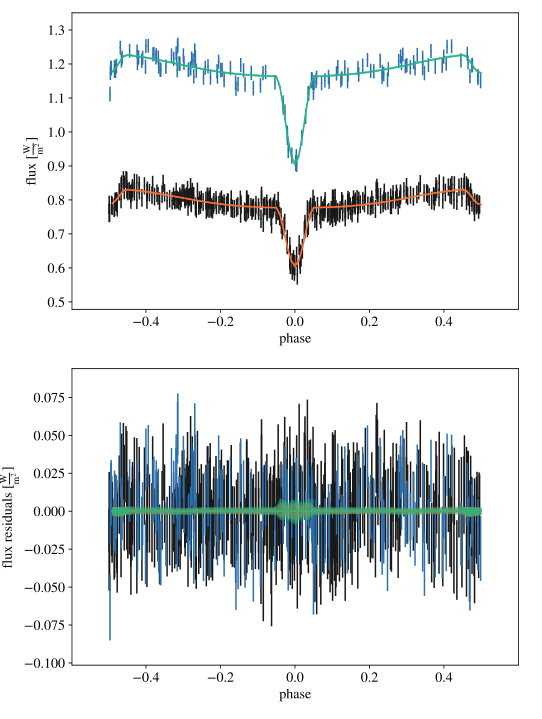}
\vspace{-0.0in}
\caption{\label{}Diagram of MCMC results and LC fitting for ZTF J190705.22+323216.9.}
\end{figure}

\begin{figure}
\centering
\vspace{-0.0in}
\includegraphics[angle=0,width=85mm]{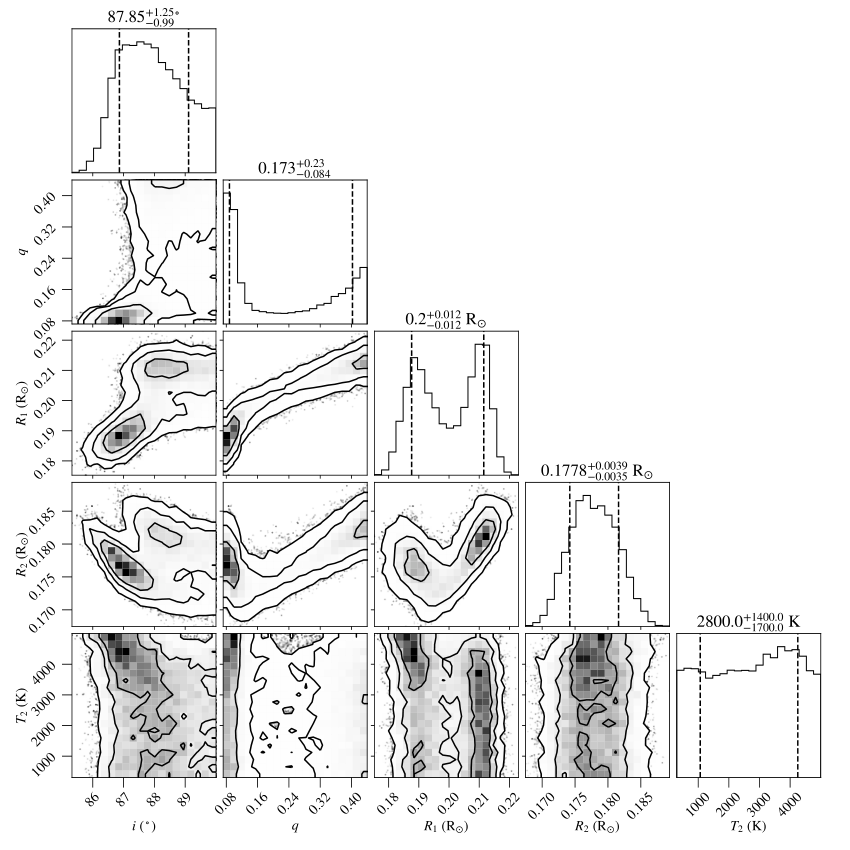}
\includegraphics[angle=0,width=64.06mm,height=85mm]{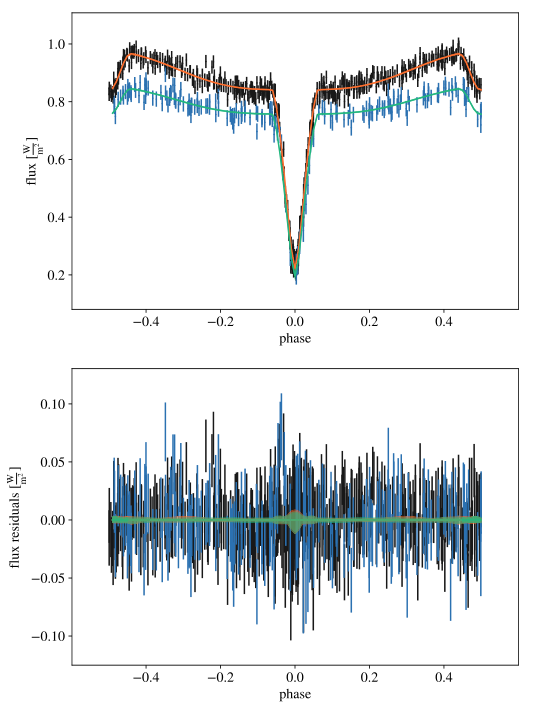}
\vspace{-0.0in}
\caption{\label{}Diagram of MCMC results and LC fitting for ZTF J192240.88+262415.5.}
\end{figure}

\begin{figure}
\centering
\vspace{-0.0in}
\includegraphics[angle=0,width=85mm]{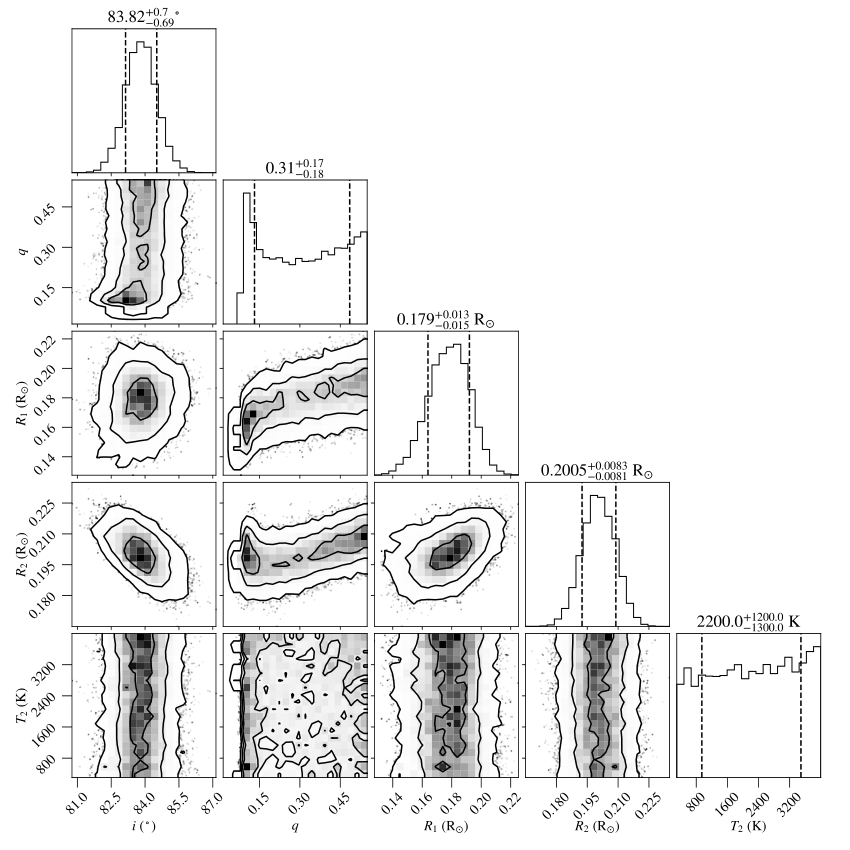}
\includegraphics[angle=0,width=64.06mm,height=85mm]{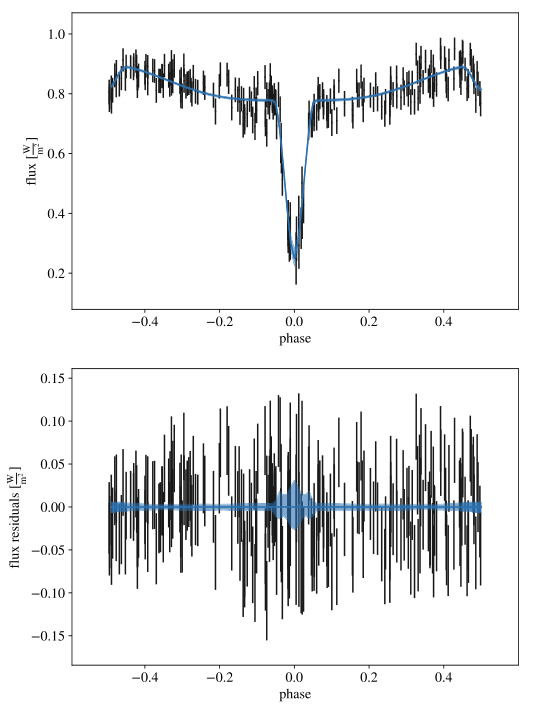}
\vspace{-0.0in}
\caption{\label{}Diagram of MCMC results and LC fitting for ZTF J192513.66+253025.6. }
\end{figure}

\begin{figure}
\centering
\vspace{-0.0in}
\includegraphics[angle=0,width=85mm]{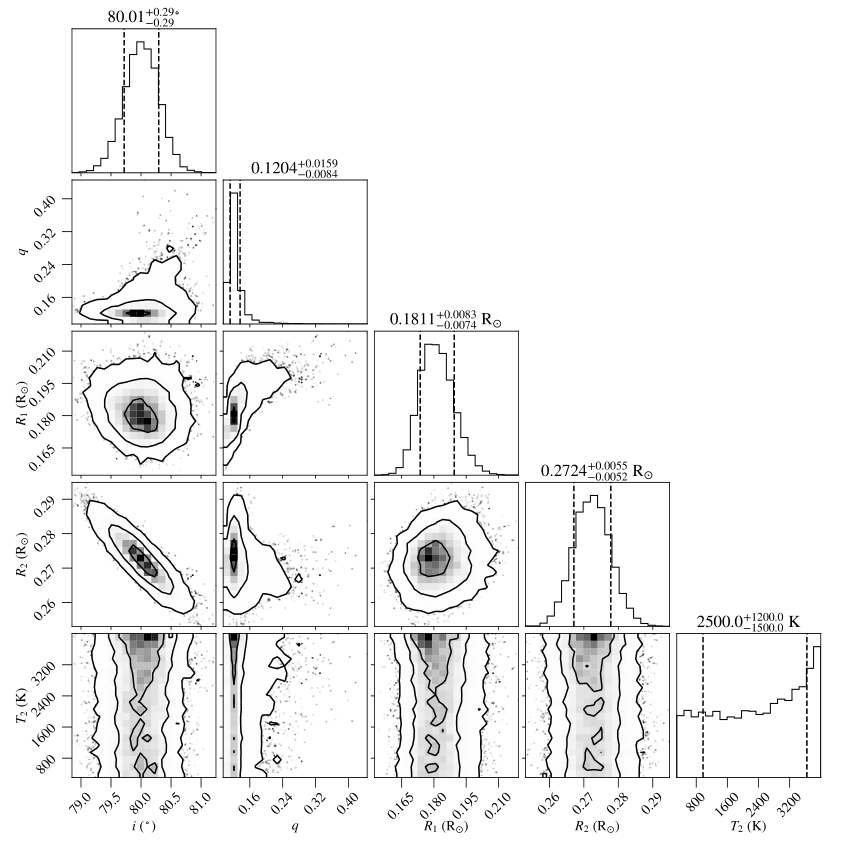}
\includegraphics[angle=0,width=64.06mm,height=85mm]{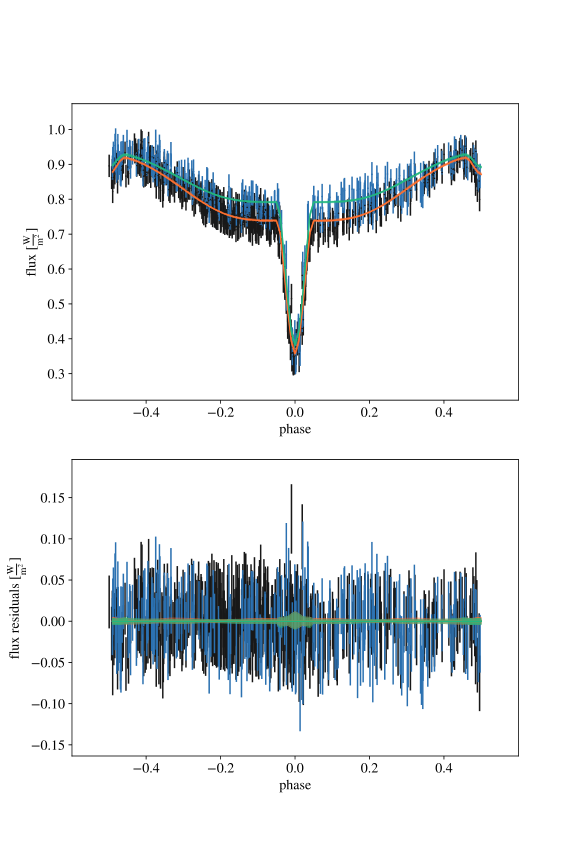}
\vspace{-0.0in}
\caption{\label{}Diagram of MCMC results and LC fitting for ZTF J193555.33+123754.8.}
\end{figure}

\begin{figure}
\centering
\vspace{-0.0in}
\includegraphics[angle=0,width=85mm]{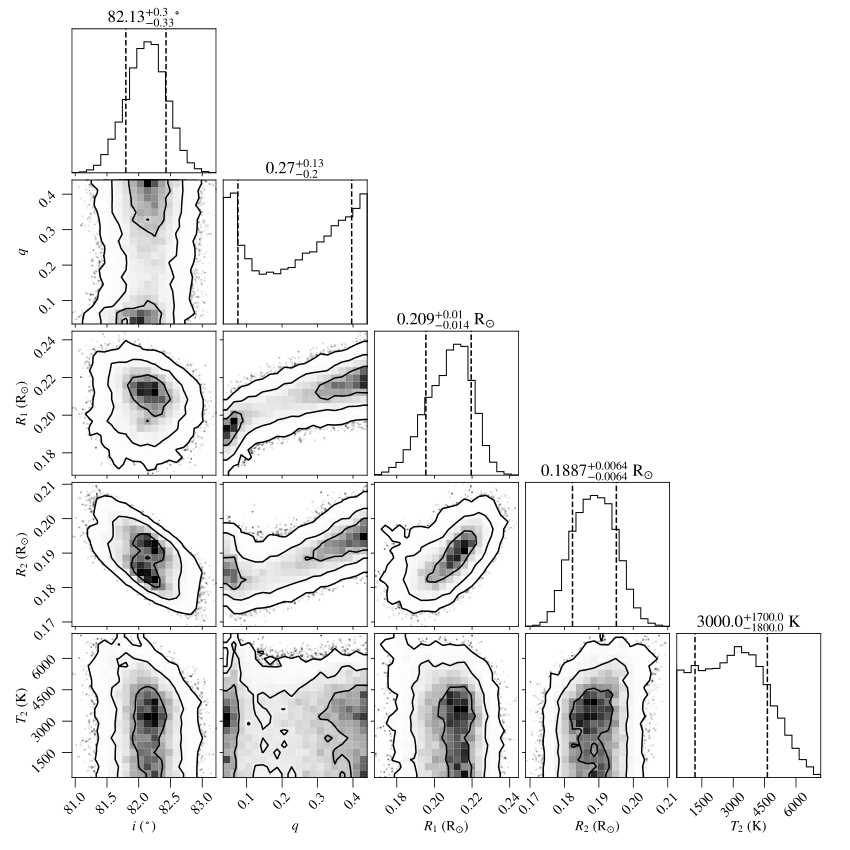}
\includegraphics[angle=0,width=64.06mm,height=85mm]{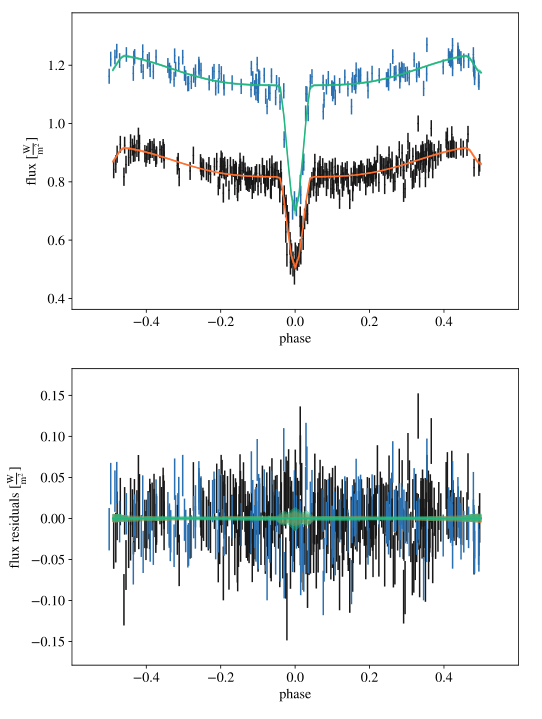}
\vspace{-0.0in}
\caption{\label{}Diagram of MCMC results and LC fitting for ZTF J193604.87+371017.2.}
\end{figure}

\begin{figure}
\centering
\vspace{-0.0in}
\includegraphics[angle=0,width=85mm]{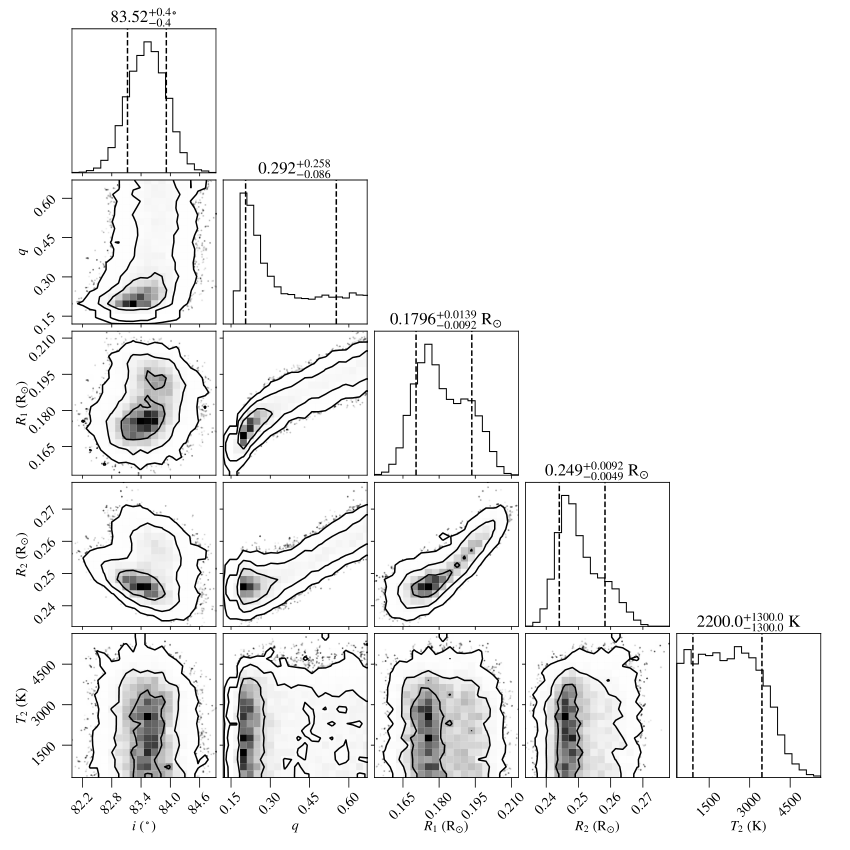}
\includegraphics[angle=0,width=64.06mm,height=85mm]{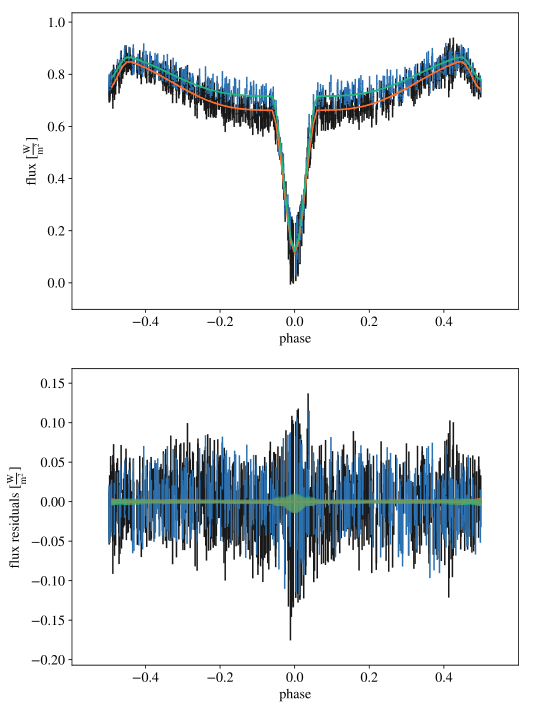}
\vspace{-0.0in}
\caption{\label{}Diagram of MCMC results and LC fitting for ZTF J193737.06+092638.7.}
\end{figure}

\begin{figure}
\centering
\vspace{-0.0in}
\includegraphics[angle=0,width=85mm]{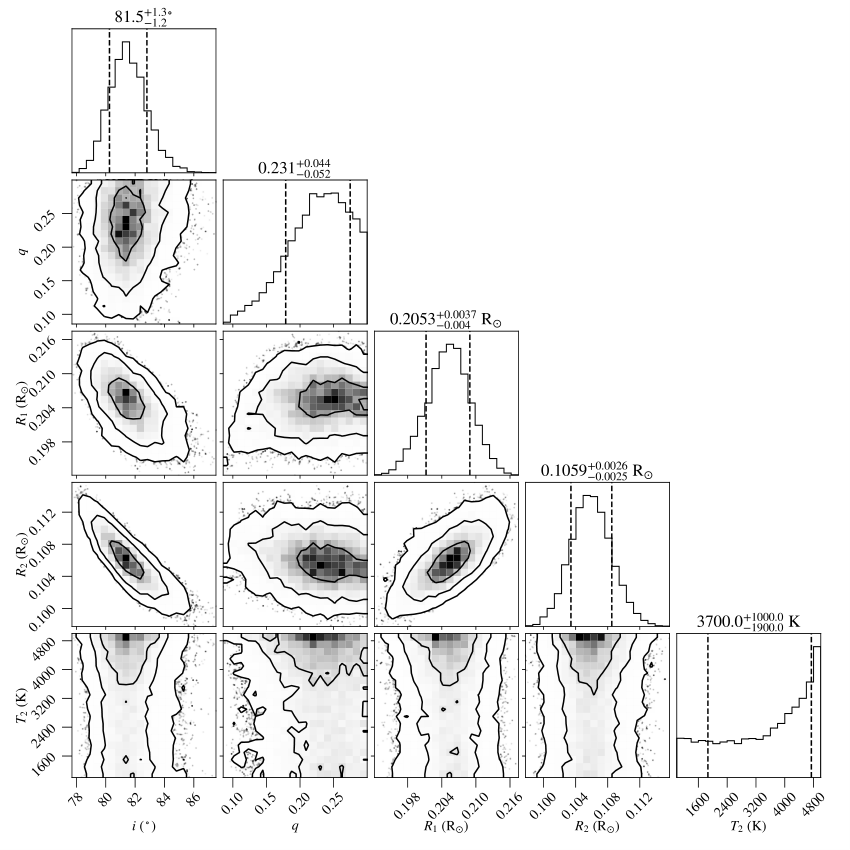}
\includegraphics[angle=0,width=64.06mm,height=85mm]{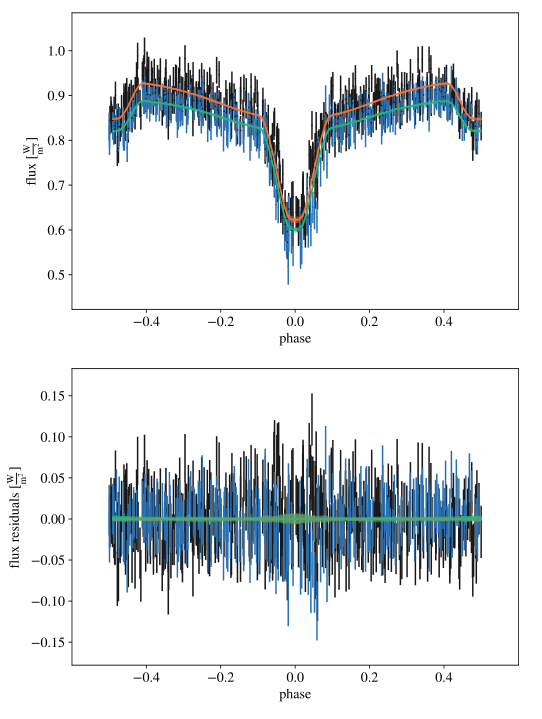}
\vspace{-0.0in}
\caption{\label{}Diagram of MCMC results and LC fitting for ZTF J195403.63+355700.6.}
\end{figure}

\begin{figure}
\centering
\vspace{-0.0in}
\includegraphics[angle=0,width=85mm]{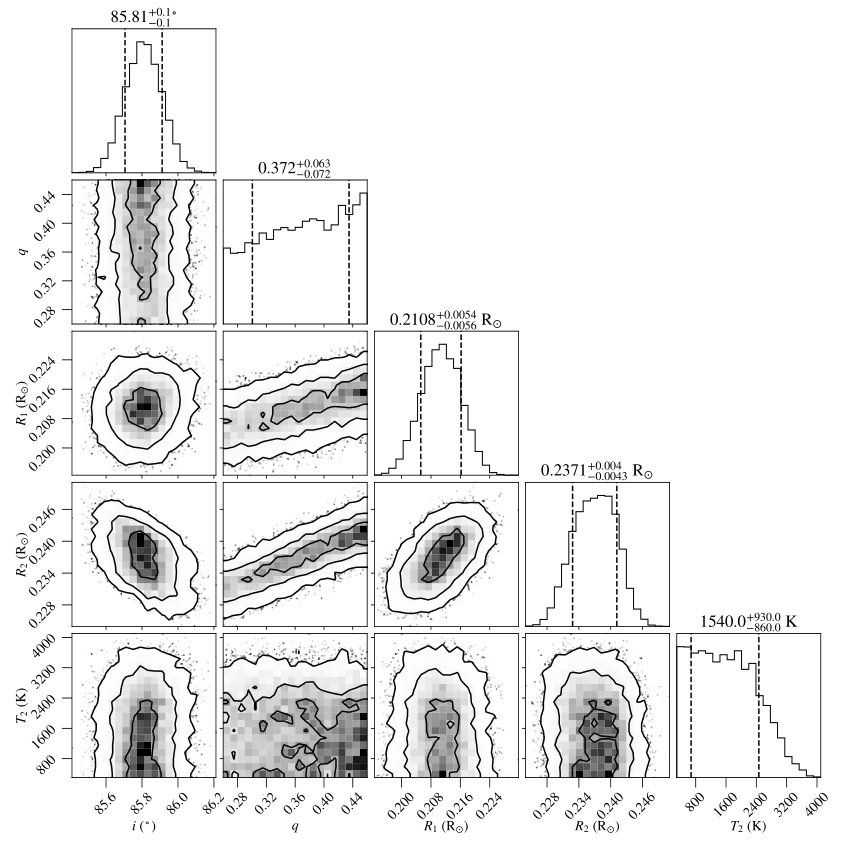}
\includegraphics[angle=0,width=64.06mm,height=85mm]{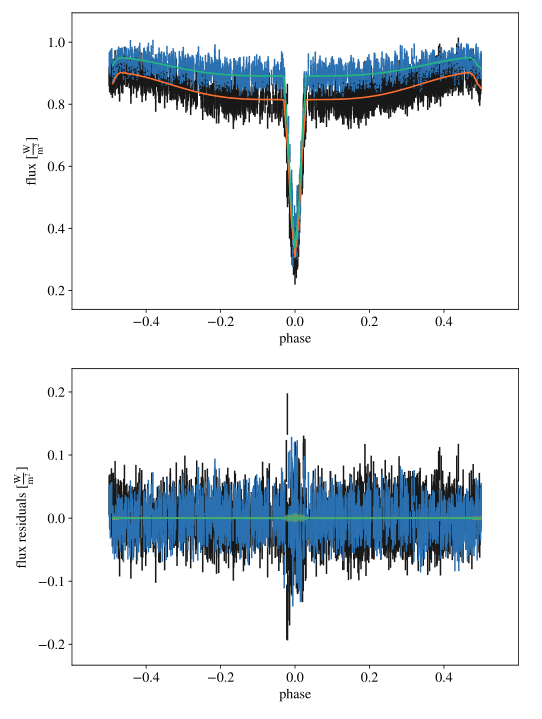}
\vspace{-0.0in}
\caption{\label{}Diagram of MCMC results and LC fitting for ZTF J195908.44+365041.1.}
\end{figure}

\begin{figure}
\centering
\vspace{-0.0in}
\includegraphics[angle=0,width=85mm]{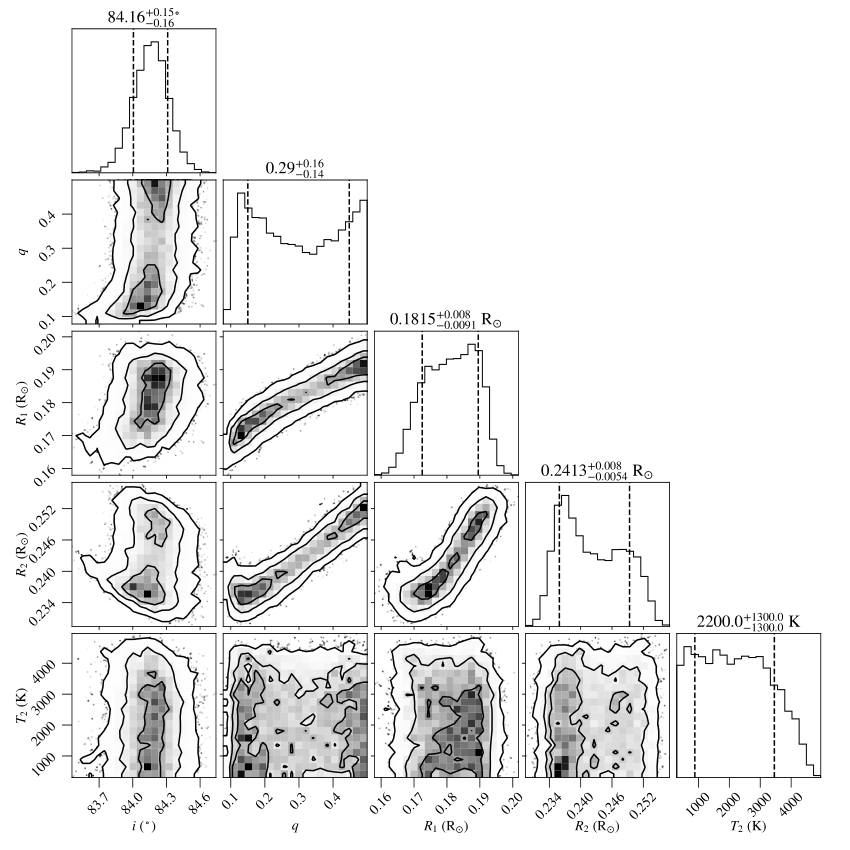}
\includegraphics[angle=0,width=64.06mm,height=85mm]{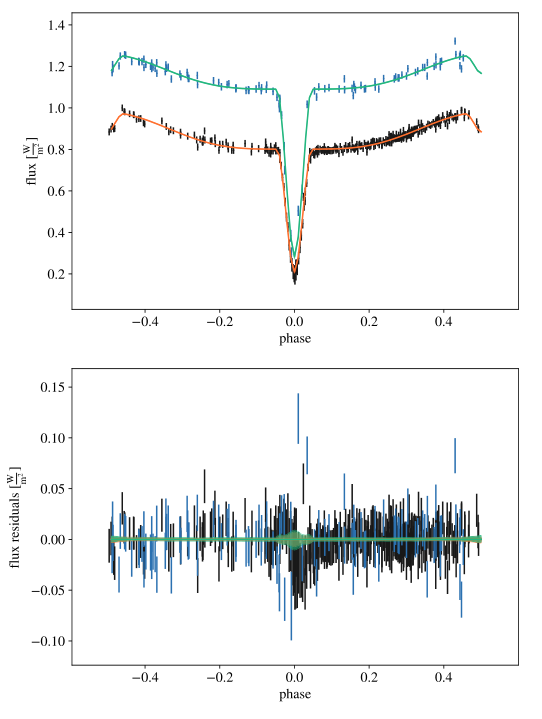}
\vspace{-0.0in}
\caption{\label{}Diagram of MCMC results and LC fitting for ZTF J200411.60+141150.2.}
\end{figure}

\begin{figure}
\centering
\vspace{-0.0in}
\includegraphics[angle=0,width=85mm]{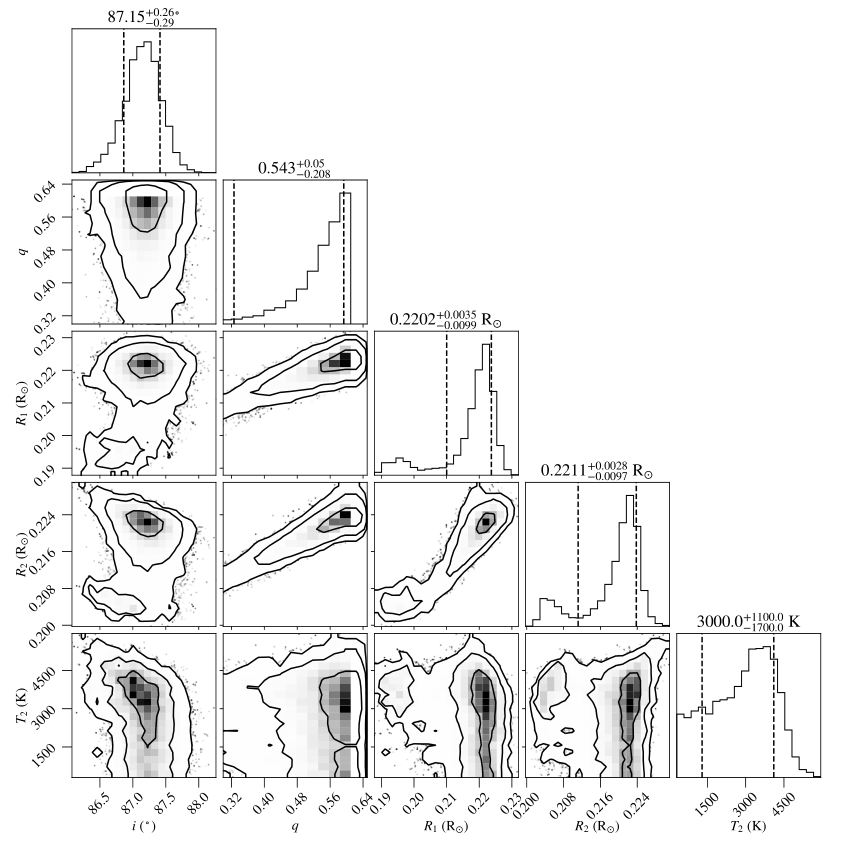}
\includegraphics[angle=0,width=64.06mm,height=85mm]{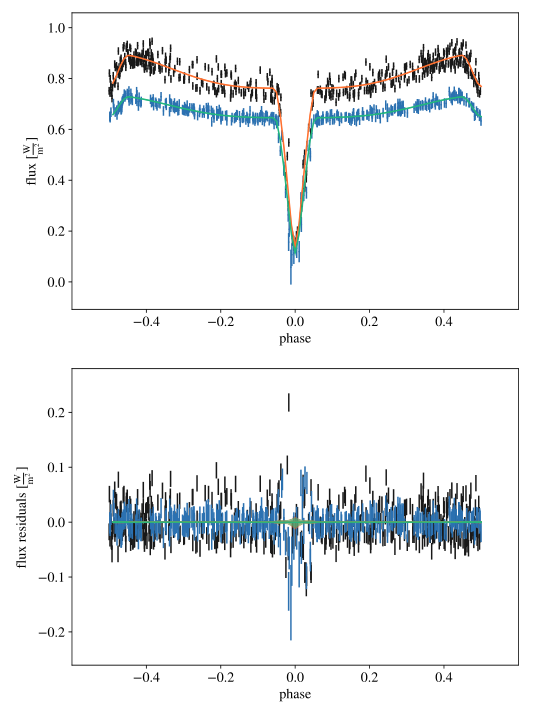}
\vspace{-0.0in}
\caption{\label{}Diagram of MCMC results and LC fitting for ZTF J203535.01+354405.0.}
\end{figure}

\begin{figure}
\centering
\vspace{-0.0in}
\includegraphics[angle=0,width=85mm]{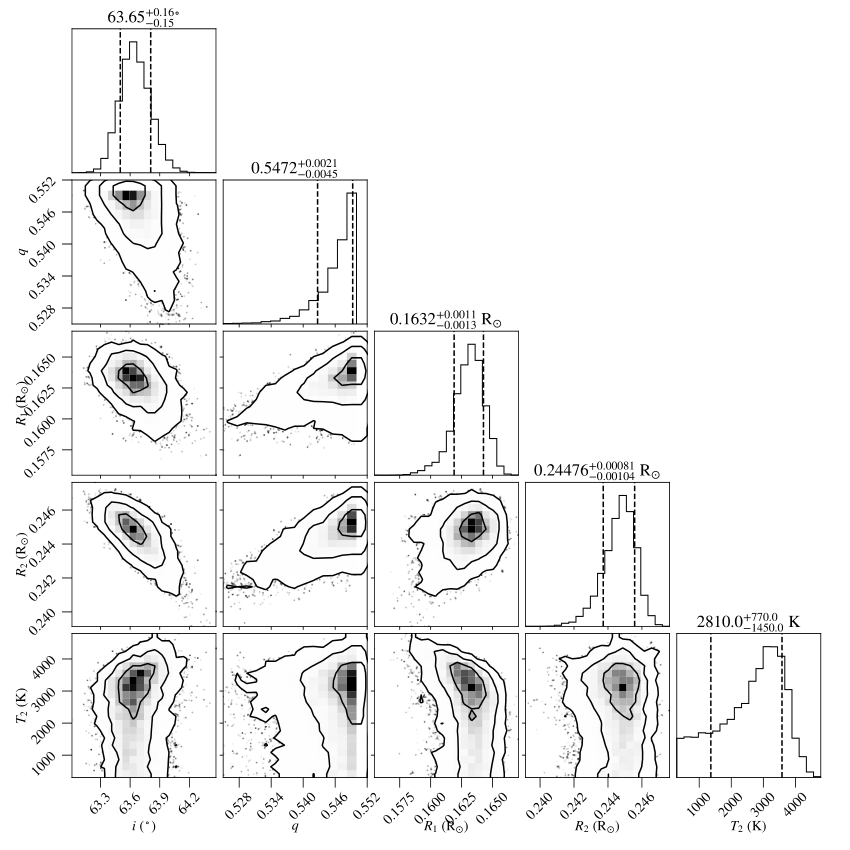}
\includegraphics[angle=0,width=64.06mm,height=85mm]{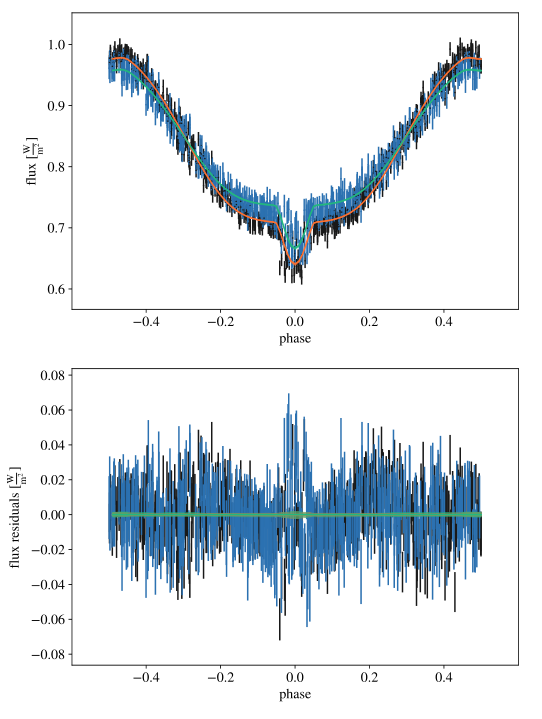}
\vspace{-0.0in}
\caption{\label{}Diagram of MCMC results and LC fitting for ZTF J204638.16+514735.5.}
\end{figure}

\begin{figure}
\centering
\vspace{-0.0in}
\includegraphics[angle=0,width=85mm]{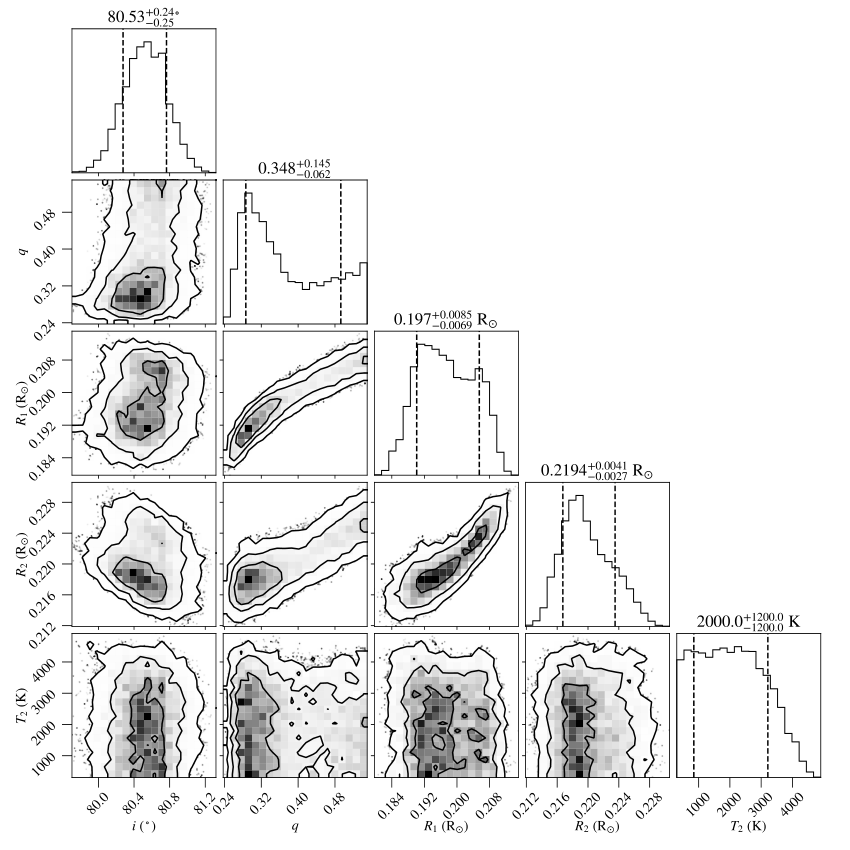}
\includegraphics[angle=0,width=64.06mm,height=85mm]{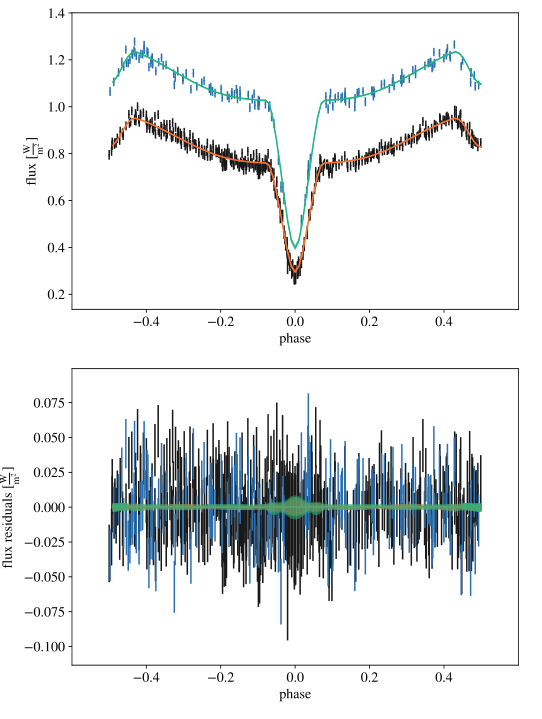}
\vspace{-0.0in}
\caption{\label{}Diagram of MCMC results and LC fitting for ZTF J210401.41+343636.3.}
\end{figure}

\begin{figure}
\centering
\vspace{-0.0in}
\includegraphics[angle=0,width=85mm]{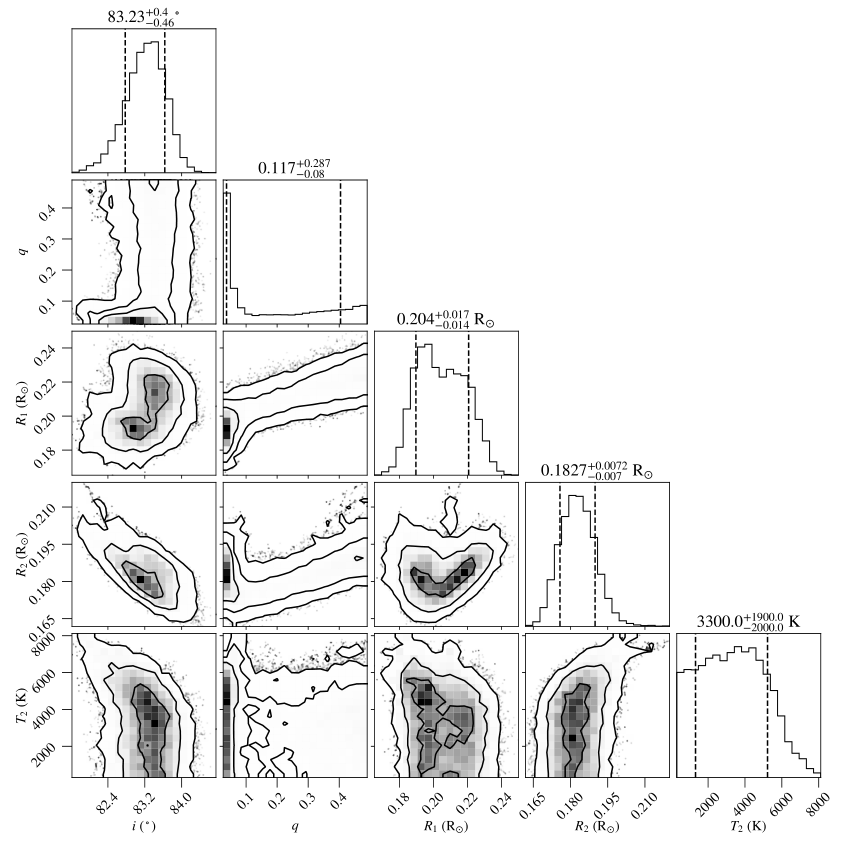}
\includegraphics[angle=0,width=64.06mm,height=85mm]{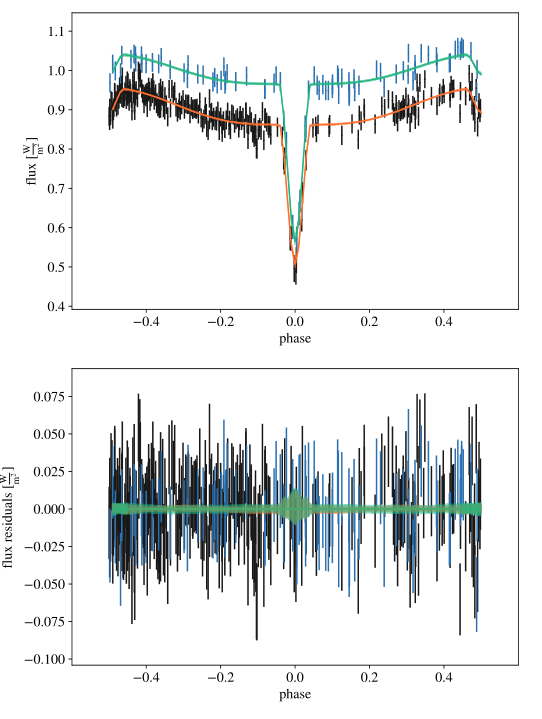}
\vspace{-0.0in}
\caption{\label{}Diagram of MCMC results and LC fitting for ZTF J221339.18+445155.8.}
\end{figure}
\begin{figure}
\centering
\vspace{-0.0in}
\includegraphics[angle=0,width=85mm]{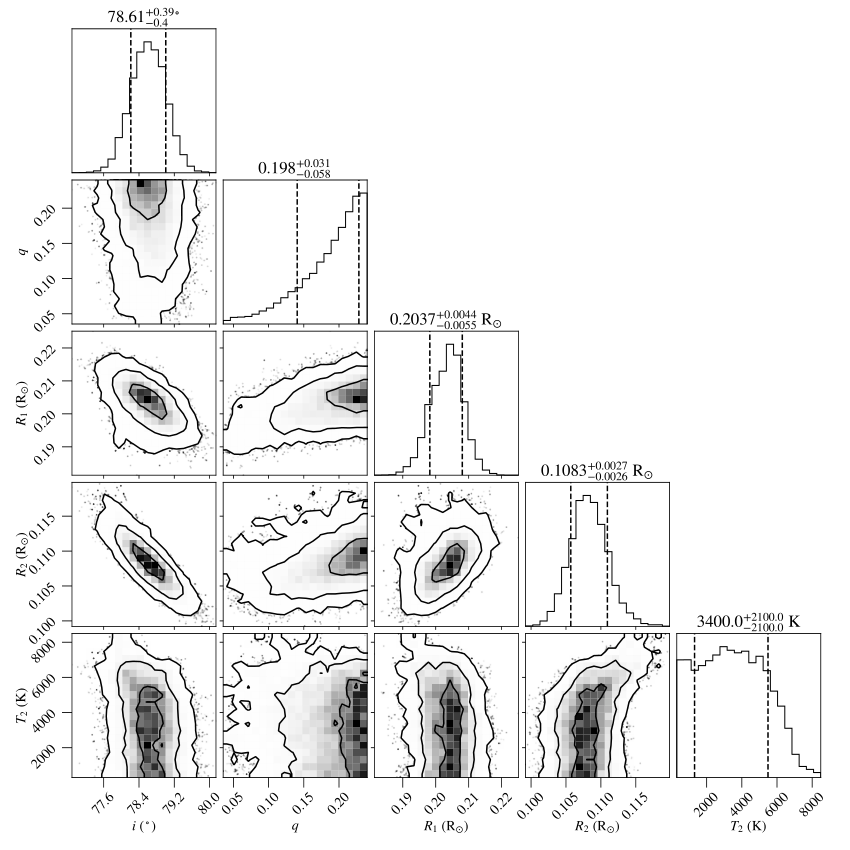}
\includegraphics[angle=0,width=64.06mm,height=85mm]{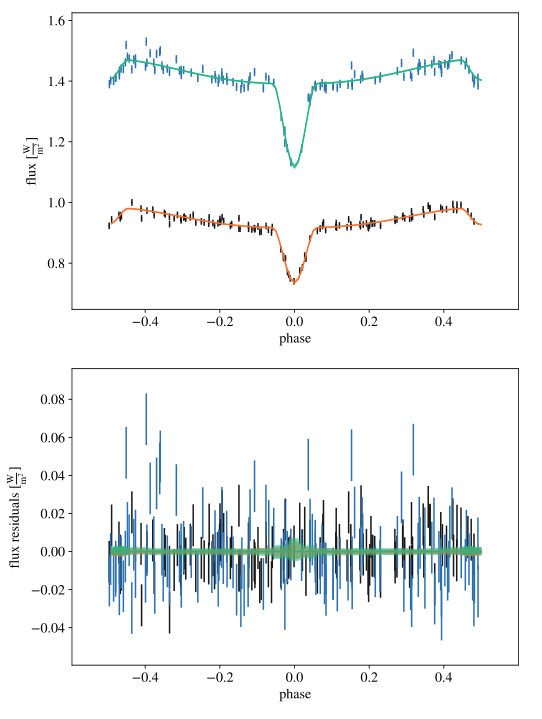}
\vspace{-0.0in}
\caption{\label{}Diagram of MCMC results and LC fitting for ZTF J223421.49+245657.1.}
\end{figure}

\label{lastpage}

\end{document}